\documentclass[a4paper,11pt,reqno]{amsart}
\usepackage[a4paper]{geometry}
\usepackage{amssymb,amsmath}
\usepackage{mathtools}
\usepackage{enumerate}
\usepackage{bookmark}
\usepackage{hyperref}
\usepackage[initials,backrefs]{amsrefs}

\numberwithin{equation}{section}
\theoremstyle{plain}
\newtheorem{theorem}{Theorem}

\newtheorem{lemma}{Lemma}

\theoremstyle{definition}
\newtheorem{definition}{Definition}
\newtheorem*{assi*}{(I) Short-range interaction}
\newtheorem*{assp*}{(P) Log-H\"older continuity condition}
\newtheorem*{dskn*}{$\dskn$}
\newtheorem*{dsknn*}{$\dsknn$}
\theoremstyle{remark}

\newcommand{\prob}[1]{\DP\left\{#1\right\}}
\newcommand{\esm}[1]{\mathbb{E}\left[\,#1\,\right]}
\newcommand{\Bone}{\mathbf{1}}
\newcommand{\BA}{\mathbf{A}}
\newcommand{\BB}{\mathbf{B}}
\newcommand{\BC}{\mathbf{C}}

\newcommand{\BG}{\mathbf{G}}
\newcommand{\BH}{\mathbf{H}}

\newcommand{\BK}{\mathbf{K}}

\newcommand{\BP}{\mathbf{P}}
\newcommand{\BU}{\mathbf{U}}

\newcommand{\BV}{\mathbf{V}}

\newcommand{\BX}{\mathbf{X}}

\newcommand{\CJ}{\mathcal{J}}

\newcommand{\CR}{\mathcal{R}}

\newcommand{\DN}{\mathbb{N}}
\newcommand{\DP}{\mathbb{P}}
\newcommand{\DR}{\mathbb{R}}
\newcommand{\DZ}{\mathbb{Z}}
\newcommand{\BDelta}{\mathbf{\Delta}}
\newcommand{\BPsi}{\mathbf{\Psi}}
\newcommand{\Bx}{\mathbf{x}}
\newcommand{\By}{\mathbf{y}}

\newcommand{\Bu}{\mathbf{u}}
\newcommand{\Bv}{\mathbf{v}}
\newcommand{\FB}{\mathfrak{B}}

\newcommand{\rA}{\mathrm{A}}
\newcommand{\rB}{\mathrm{B}}

\newcommand{\rR}{\mathrm{R}}
\newcommand{\rS}{\mathrm{S}}

\DeclareMathOperator{\card}{card}
\DeclareMathOperator{\diam}{diam}
\DeclareMathOperator{\dist}{dist}
\DeclareMathOperator{\supp}{supp}

\newcommand{\ee}{\mathrm{e}}

\newcommand{\comp}{\mathrm{c}}
\newcommand{\fui}{\mathrm{FI}}

\newcommand{\pai}{\mathrm{PI}}
\newcommand{\sep}{\mathrm{sep}}

\newcommand{\condI}{\mathbf{(I)}}
\newcommand{\condP}{\mathbf{(P)}}
\newcommand{\dsk}[1]{\mathbf{(DS.}k,#1,N\mathbf{)}}
\newcommand{\dsn}[2]{\mathbf{(DS.}#1#2,$\,N\mathbf{)}$}

\newcommand{\dskn}{\mathbf{(DS.}k,N\mathbf{)}}
\newcommand{\dsknn}{\mathbf{(DS.}k,n,N\mathbf{)}}

\newcommand{\dskonn}{\mathbf{(DS.}k+1,n,N\mathbf{)}}
\newcommand{\tto}[1]{\smash{\mathop{\,\,\,\, \longrightarrow \,\,\,\, }\limits_{#1}}}

\begin{document}
\title[localization for weakly interacting continuous models]{Anderson localization for weakly interacting multi-particle models in the continuum}

\author[T.~Ekanga]{Tr\'esor EKANGA$^{\ast}$}

\address{$^{\ast}$%
Institut de Math\'ematiques de Jussieu,
Universit\'e Paris Diderot,
Batiment Sophie Germain,
13 rue Albert Einstein,
75013 Paris,
France}
\email{tresor.ekanga@imj-prg.fr}
\date{\today}
\begin{abstract}
For the weakly interacting one-dimensional multi-particle Anderson  model in the continuum space of configurations, we prove the spectral exponential and the strong dynamical localization. The results require the interaction amplitude to be sufficiently small. The general strategy uses the multi-scale analysis bounds. Actually, we show that the multi-scale analysis bounds of the single particle model remain stable when passing to multi-particle systems, provided that the inter-particle interaction is sufficiently small. The common probability distribution of the i.i.d. random external potential in the Anderson model, is only needed to be log-H\"older continuous.
\end{abstract}
\maketitle 

\section{Introduction}

The analysis of multi-particle quantum systems is relatively recent and there is a number of results in both the discrete and the continuum cases see for example \cites{AW09,CS09b,FW15,KN13,E11,E12,E13}. In the papers \cites{AW09,AW10,BCSS10a,C11,C12,CS08,CS09b,KN13}, the authors analyzed multi-particle systems and  proved the Anderson localization in the high disorder limit. For the localization results in the single-particle theory, we refer to \cites{CL90,FMSS85,K08,GB98,GK01}. Aizenmann and Warzel \cite{AW09}, used the adaptation to multi-particle systems of the fractional moment method. Chulaevsky and Suhov their-selves developed for the strong disorder regime the same extension for the multi-scale analysis. While, in \cite{KN13}, Klein and Nguyen, extended to multi-particle systems the so-called bootstrap multi-scale analysis. All these works were done under the strong disorder regime. We adapted in our papers \cites{E11,E12}, the multi-scale analysis to multi-particle systems under the low energy regime.

We are concerned in this work to the weakly interacting regime of the multi-particle Anderson model in the continuum. Localization in that case was obatined by Aizenmann and Warzel \cite{AW09}. But the exponential decay of the eigenfunctions was established in the hausdorff distance and due to technical requirements of the fractional moment method they assumed  the common probability distribution of the i.i.d. random external potential to be absolutely continuous with a bounded density. In our work \cite{E13}, we showed the exponential decay of the eigenfunctions in the max-norm under a weaker assumption of log-H\"{o}lder continuity of the common probability distribution function via the multi-particle multi-scale analysis. The method is based on  the continuum version of the multi-particle multi-scale analysis which is exposed for single-particle models in the book by Stollmann \cite{St01}. Note that in \cite{FW15}, the authors extended the multi-particle fractional moment method to the continuous space and obtained in the same occasion, localization for multi-particle systems.

For single-particle models in one dimension, the complete Anderson localization occurs even for singular probability distributions, such as Bernoulli's measures \cites{CKM87,DSS02}. We prove here that it also remains true for weakly interacting  multi-particle quantum systems with log-H\"older distributions functions. This last assumption is important for the Wegner estimates which are used in the multi-scale analysis scheme. The scheme developed in \cite{E13} in the discrete case, has been modified because the spectrum of a compact and self-adjoint  operator is not necessary finite. This problem is resolved in our case with the help of  the Weyl's law. Our main results are Theorem 1 (spectral exponential localization) and Theorem 2 (strong dynamical localization).

We now discuss on the structure of the paper. The rest of this section is devoted to the model, assumptions and the mains results on the complete Anderson localization with weak interaction. In section 2, we describe the multi-scale analysis scheme ( geometric facts and probability bounds useful for the MSA). We prove in section 3, the initials bounds of the multi-particle multi-scale analysis. We develop in section 4, the multi-scale induction step of the multi-scale analysis. In section 5, the multi-particle multi-scale analysis estimates are proved at all length scales and for any number of particles less or equal to the total number of particles (which is assumed to be finite).   

\subsection{The model}
We fix at the very beginning the number of particles $N\geq 2$. We are concern with multi-particle random Schr\"odinger operators of the following forms:
\[
\BH^{(N)}(\omega):=-\BDelta + \BU+\BV,
\]
acting in $L^{2}((\DR^{d})^N)$. Sometimes, we will use the identification $(\DR^{d})^N\cong \DR^{Nd}$. Above, $\BDelta$ is the Laplacian on $\DR^{Nd}$, $\BU$ represents the inter-particle interaction which acts as multiplication operator in $L^{2}(\DR^{Nd})$. Additional information on $\BU$ is given in the assumptions. $\BV$ is the multi-particle random external potential also acting as multiplication operator on $L^{2}(\DR^{Nd})$. For $\Bx=(x_1,\ldots,x_N)\in(\DR^{d})^N$, $\BV(\Bx)=V(x_1)+\cdots+ V(x_N)$ and $\{V(x,\omega), x\in\DR^d\}$ is a random i.i.d. stochastic process relative to some probability space $(\Omega,\FB,\DP)$.

Observe that the non-interacting Hamiltonian $\BH^{(N)}_0(\omega)$ can be written as a tensor product:
\[
\BH^{(N)}_0(\omega):=-\BDelta +\BV=\sum_{k=1}^N \Bone^{\otimes(k-1)}_{L^{2}(\DR^d)}\otimes H^{(1)}(\omega)\otimes \Bone^{\otimes(N-k)}_{L^2(\DR^d)},
\]
where, $H^{(1)}(\omega)=-\Delta + V(x,\omega)$ acting on $L^2(\DR^d)$. We will also consider random Hamiltonian $\BH^{(n)}(\omega)$, $n=1,\ldots,N$ defined similarly. Denote by $|\cdot|$ the max-norm in $\DR^{nd}$.

\subsection{Assumptions}

\begin{assi*}

Fix any $n=1,\ldots,N$. The potential of inter-particle interaction $\mathbf{U}$ is bounded and of the form
\[
\BU(\Bx)=\sum_{1\leq i<j\leq n}\Phi(|x_i-x_j|),\quad \Bx=(x_1,\ldots,x_n),
\]
where  $\Phi:\DN:\rightarrow\DR$ is a compactly supported function such that

\begin{equation}\label{eq:finite.range.k}
\exists r_0\in\DN: \supp \Phi\subset[0,r_0].
\end{equation}

\end{assi*}

Set $\Omega=\DR^{\DZ^d}$ and $\FB=\bigotimes_{\DZ^d}\mathcal{B}(\DR)$ where $\mathcal{B}(\DR)$ is the Borel sigma-algebra on $\DR$. Let $\mu$ be a probability measure on $\DR$ and define $\DP=\bigotimes_{\DZ^d}\mu$ on $\Omega$.

The external random potential $V\colon\DZ^d\times\Omega\to\DR$ is an i.i.d. random field
relative to  $(\Omega,\FB,\DP)$ and is defined by $V(x,\omega)=\omega_x$ for $\omega=(\omega_i)_{i\in\DZ^d}$.
The common probability distribution function, $F_V$,  of the i.i.d. random variables $V(x,\cdot)$, $x\in\DZ^d$ associated to the measure $\mu$ is defined by
\[
F_V: t \mapsto \prob{V(0,\omega)\leq t }.
\]
\begin{assp*}
The random potential $V:\DZ^d\times\Omega\rightarrow \DR$ is i.i.d. and the corresponding probability distribution function $F_V$ is log-H\"older continuous:  More precisely,
\begin{align}\label{eq:assumption.3prime}
&s(F_V,\varepsilon) := \sup_{a\in\DR}(F_V(a+\varepsilon)-F_V(a))
\leq\frac{C}{|\ln\epsilon|^{2A}}\\
&\text{for some }C\in(0,\infty)\text{ and }A>\frac{3}{2}\times4^Np+9Nd.\notag
\end{align}
Note that this last condition depends on the parameter $p$ which will be introduced  in Section \ref{sec:Nparticle.scheme}.
\end{assp*}

\subsection{The results}\label{sec:main.results}

\begin{theorem}\label{thm:exp.loc}
Let $d=1$. Under assumptions $\condI$ and $\condP$,
 there exists $h^*>0$ such that for any $h\in(-h^*, h^*)$ the Hamiltonian $\BH^{(N)}_h$, with interaction of amplitude $|h|$, exhibits complete Anderson localization, \emph{i.e.}, with $\DP$-probability one,
the spectrum of $\BH^{(N)}_h$  is pure point, and each eigenfunction $\BPsi$ is exponentially decaying at infinity:
\[
\|\chi_{\Bx}\cdot\BPsi\|\leq C\ee^{-c|\Bx|},
\]
for some positive constants $c$ and $C$.
\end{theorem}

\begin{theorem}\label{thm:dynamical.loc}
Under assumptions $\condI$ and  $\condP$, there exist $h^*>0$, $s^*>0$ such that for any $h\in(-h^*,h^*)$, any $s\in(0,s^*)$ and any compact domain $\BK\subset \DR^{Nd}$, we have:
\[
\esm{\sup_{t>0}\| \BX^{s}\ee^{-it\BH^{(N)}(\omega)}\BP_I(\BH^{(N)}(\omega))\Bone_{\BK}\|_{L^2(\DR^{Nd})}}<\infty,
\]
where $(|\BX|\BPsi)(\Bx):=|\Bx|\BPsi(\Bx)$, $\BP_{I}(\BH^{(N)}(\omega))$ is the spectral projection of $\BH^{(N)}(\omega)$ onto the interval $I$ and $\Bone_{\BK}$ is characteristic function of the set $\BK$.
\end{theorem}

\section{The multi-particle multi-scale analysis scheme}\label{sec:Nparticle.scheme}

\subsection{Geometric facts}
According to the general structure of the MSA, we work with  \emph{rectangular} domains. For $\Bu=(u_1,\ldots,u_n)\in\DZ^{nd}$, we denote by $\BC^{(n)}_L(\Bu)$ the $n$-particle open cube, i.e,
\[
\BC^{(n)}_L(\Bu)=\left\{\Bx\in\DR^{nd}:|\Bx-\Bu|< L\right\},
\]
and given $\{L_i: i=1,\ldots,n\}$, we define the rectangle
\begin{equation}          \label{eq:cube}
\BC^{(n)}(\Bu)=\prod_{i=1}^n C^{(1)}_{L_i}(u_i),
\end{equation}
where $C^{(1)}_{L_i}(u_i)$ are cubes of side length $2L_i$ center at points $u_i\in\DZ^d$. We also define 
\[
\BC^{(n,int)}_L(\Bu):=\BC^{(n)}_{L/3}(\Bu), \quad \BC^{(n,out)}_L(\Bu):=\BC^{(n)}_L(\Bu)\setminus\BC^{(n)}_{L-2}(\Bu), \quad \Bu\in\DZ^{nd}
\]
and introduce  the characteristic functions:
\[
\Bone^{(n,int)}_{\Bx}:=\Bone_{\BC^{(n,int)}_L(\Bx)}, \qquad \Bone^{(n,out)}_{\Bx}:= \Bone_{\BC^{(n,out)}_L(\Bx)}.
\]
The volume of the cube $\BC^{(n)}_L(\Bu)$ is $|\BC_L^{(n)}(\Bu)| :=(2L)^{nd}$.
We denote the restriction of the Hamiltonian $\BH_h^{(n)}$ to  $\BC^{(n)}(\Bu)$ by
\begin{align*}
&\BH_{\BC^{(n)}(\Bu),h}^{(n)}=\BH^{(n)}_h\big\vert_{\BC^{(n)}(\Bu)}\\
&\text{with Dirichlet boundary conditions}
\end{align*}

We denote the spectrum of $\BH_{\BC^{(n)}(\Bu),h}^{(n)}$  by
$\sigma\bigl(\BH_{\BC^{(n)}(\Bu)}^{(n),h}\bigr)$ and its resolvent by
\begin{equation}\label{eq:def.resolvent}
\BG^{(n)}_{\BC^{(n)}(\Bu),h}(E):=\Bigl(\BH_{\BC^{(n)}(\Bu),h}^{(n)}-E\Bigr)^{-1},\quad E\in\DR\setminus\sigma\Bigl(\BH_{\BC^{(n)}(\Bu),h}^{(n)}\Bigr).
\end{equation}

Let $m>0$ and $E\in\DR$ be given.  A cube $\BC_L^{(n)}(\Bu)\subset\DR^{nd}$, $1\leq n\leq N$ will be  called $(E,m,h)$-\emph{nonsingular} ($(E,m,h)$-NS) if $E\notin\sigma(\BH^{(n)}_{\BC^{(n)}_{L}(\Bu),h})$ and
\begin{equation}\label{eq:singular}
\|\Bone^{(n,out)}_{\Bx}\BG^{(n)}_{\BC^{(n)}_L(\Bx)}(E)\Bone^{(n,int)}_{\Bx}\|\leq\ee^{-\gamma(m,L,n)L},
\end{equation}
where
\begin{equation}\label{eq:gamma}
\gamma(m,L,n)=m(1+L^{-1/8})^{N-n+1}.                     
\end{equation}
Otherwise it will be called $(E,m,h)$-\emph{singular} ($(E,m,h)$-S).

Let us introduce the following.
\begin{definition}
Let $n\geq 1$, $E\in\DR$ and $\alpha=3/2$.  
\begin{enumerate}[\rm(A)]
\item
A cube $\BC_L^{(n)}(\Bv)\subset\DR^{nd}$  is called $(E,h)$-resonant ($(E,h)$-R) if
\begin{equation} \label{eq:E-resonant}
\dist\Bigl[E,\sigma\bigl(\BH_{\BC_L^{(n)}(\Bv),h}^{(n)}\bigr)\Bigr]\leq\ee^{-L^{1/2}}.
\end{equation}
Otherwise it is called $(E,h)$-non-resonant ($(E,h)$-NR).
\item
A cube $\BC_L^{(n)}(\Bv)\subset\DR^{nd}$ is called $(E,h)$-completely nonresonant ($(E,h)$-CNR), if it does not contain any $(E,h)$-R cube of size $\geq L^{1/\alpha}$. In particular $\BC^{(n)}_L(\Bv)$ is itself $(E,h)$-NR.
\end{enumerate}
\end{definition}

We will also make use of the following notion.

\begin{definition}\label{def:separability}
A cube $\BC^{(n)}_L(\Bx)$ is  $\CJ$-separable from $\BC^{(n)}_L(\By)$ if there exists a nonempty subset $\CJ\subset\{1,\cdots,n\}$ such that
\[
\left(\bigcup_{j\in \CJ}C^{(1)}_{L}(x_j)\right)\cap
\left(\bigcup_{j\notin \CJ}C_L^{(1)}(x_j)\cup \bigcup_{j=1}^n C_L^{(1)}(y_j)\right)=\emptyset.
\]
A pair $(\BC^{(n)}_L(\Bx),\BC^{(n)}_L(\By))$ is separable if $|\Bx-\By|>7NL$ and if one of the cube is $\CJ$-separable from the other.
\end{definition}
\begin{lemma}\label{lem:separable.distant}
Let $L>1$.
\begin{enumerate}[\rm(A)]
\item
For any $\Bx\in\DZ^{nd}$, there exists a collection of $n$-particle cubes
$\BC^{(n)}_{2nL}(\Bx^{(\ell)})$ with $\ell=1,\ldots,\kappa(n)$, $\kappa(n)= n^n$, $\Bx^{\ell}\in\DZ^{nd}$ such that if
 $\By\in\DZ^{nd}$ satisfies $|\By-\Bx|>7NL$ and
\[
\By \notin \bigcup_{\ell=1}^{\kappa(n)} \BC^{(n)}_{2nL}(\Bx^{(\ell)})
\]
then the cubes $\BC^{(n)}_L(\Bx)$ and $\BC^{(n)}_L(\By)$ are separable.
\item
Let $\BC^{(n)}_L(\By)\subset \DR^{nd}$ be an $n$-particle cube. Any cube  $\BC^{(n)}_L(\Bx)$ with 
\[
|\By-\Bx|>\max_{1\leq i,j\leq n}|y_i-y_j| +5NL,
\]
 is $\CJ$-separable from
$\BC^{(n)}_L(\By)$ for some $\CJ\subset\{1,\ldots,n\}$.
\end{enumerate}
\end{lemma}
\begin{proof}
See the appendix section \ref{sec:appendix}.
\end{proof}

\subsection{The multi-particle Wegner estimates}
In our earlier work \cite{E12}  as well as in other previous papers in the multi-particle localization theory \cites{CS09b,BCSS10b} the notion of separability was crucial in order to prove the Wegner estimates for pairs of multi-particle cubes via the Stollmann's Lemma. It is plain (cf. \cite{E12}, Section 4.1), that  sufficiently distant pairs of fully interactive cubes have disjoint projections and this fact combined with independence is used in that case to bound the probability of an intersection of events relative to those projections. We state below the Wegner estimates directly in a form suitable for our multi-particle multi-scale analysis using assumption $\condP$.

\begin{theorem}\label{thm:Wegner}
 Assume that the random potential satisfies assumption $\condP$, then 
\begin{enumerate}

\item[\rm(A)]
for any $E\in\DR$
\begin{equation}\label{eq:cor.Wegner.2A}
\prob{\text{$ \BC^{(n)}_L(\Bx)$ is not $E$-CNR }}\leq L^{-p\,4^{N-n}},
\end{equation}

\item[\rm(B)]
\begin{equation}\label{eq:cor.Wegner.2B}
\prob{\text{$\exists E\in \DR:$ neither $\BC^{(n)}_{L}(\Bx)$ nor $\BC^{(n)}_{L}(\By)$ is $E$-CNR}} \leq L^{-p\,4^{N-n}},
\end{equation}
\end{enumerate}
where $p>6Nd$, depends only on the fixed number of particles $N$ and the configuration dimension $d$.
\end{theorem}

\begin{proof}
See the article \cite{BCSS10b}.
\end{proof}

We recall below the geometric resolvent inequality and the eigenfunction decay inequality.

\begin{theorem}[Geometric resolvent inequality (GRI)]\label{thm:GRI.GF}
For a given bounded interval $I_0\subset\DR$, there is a constant $C_{geom}>0$ such that for $\BC^{(n)}_{\ell}(\Bx)\subset\BC^{(n)}_L(\Bu)$, $\BA\subset\BC^{(n,int)}_{\ell}(\Bx)$, $\BB\subset\BC^{(n)}_L(\Bu)\setminus\BC^{(n)}_{\ell}(\Bx)$ and $E\in I_0$, the following inequality holds true:
\[
\|\Bone_{\BB}\BG^{(n)}_{\BC^{(n)}_L(\Bu)}(E)\Bone_{\BA}\|\leq C_{geom}\cdot\|\Bone_{\BB}\BG^{(n)}_{\BC^{(n)}_L(\Bu)}(E)\Bone_{\BC^{(n,out)}_{\ell}(\Bx)}\|\cdot\|\Bone_{\BC^{(n,out)}_{\ell}(\Bx)}\BG^{(n)}_{\BC^{(n)}_{\ell}(\Bx)}(E)\Bone_{\BA}\|.
\]
\end{theorem}
\begin{proof}
See \cite{St01}, Lemma 2.5.4.
\end{proof}

\begin{theorem}[Eigenfunctions decay inequality (EDI)] \label{thm:GRI.EF}
For every $E\in\DR$, $\BC^{(n)}_{\ell}(\Bx)\subset \DR^{nd}$ and every polynomially bounded function $\BPsi\in L^2(\DR^{nd})$:
\[
\|\Bone_{\BC^{(n)}_1(\Bx)}\cdot\BPsi\|\leq C\cdot\|\Bone_{\BC^{(n,out)}_{\ell}(\Bx)}\BG^{(n)}_{\BC^{(n)}_{\ell}(\Bx)}(E)\Bone_{\BC^{(n,int)}_{\ell}(\Bx)}\|\cdot\|\Bone_{\BC^{(n,out)}_{\ell}(\Bx)}\cdot\BPsi\|.
\]
\end{theorem}
\begin{proof}
See section 2.5 and proposition 3.3.1 in \cite{St01}.
\end{proof}

\section{The initial MSA bound for the weakly interacting multi-particle system}

\subsection{The fixed energy  MSA bound for the n-particle system without interaction}\label{sec:non.int.initial.bound}

We begin with the well known single-particle exponential localization for the eigenfunctions and for one-dimensional Anderson models in the continuum proved in the paper by Damanik et al. \cite{DSS02}. Let $H^{(1)}_{C^{(1)}_L(x)}(\omega)$ be the restriction of the single-particle Hamiltonian into the cube $C^{(1)}_L(x)$ and denote by $\{\lambda_j:\varphi_j\}_{j\geq 0}$ its eigenvalues and corresponding eigenfunctions. We have the following, namely the single-particle exponential localization for the eigenfunctions in any cube.

\begin{theorem}[Single-particle localization]\label{thm:1p.loc}
There exists a constant $\widetilde{\mu}>0$ such that for every generalized eigenfunction $\varphi$ of the single-particle Hamiltonian $H^{(1)}_{C^{(1)}_L(x)}(\omega)$, we have:
\[
\esm{\left\|\Bone_{C^{(1,out)}_{L}(x)}\cdot\varphi\cdot\Bone_{C^{(1,int)}_L(x)}\right\|}\leq \ee^{-\widetilde{\mu}L}.
\]
\end{theorem}
\begin{proof}
We refer to the book by Stollmann \cite{St01}.
\end{proof}

The main result of this subsection is Theorem \ref{thm:initial.bound.np} given below.
The proof of Theorem \ref{thm:initial.bound.np} relies on an auxiliary statement, Lemma \ref{lem:1p.NS.implies.np.NS}. We need to introduce first
\[
\{(\lambda^{(i)}_{j_i},\psi^{(i)}_{j_i}):j_i\geq 1\},                
\]
the eigenvalues and the corresponding eigenfunctions of $H^{(1)}_{C^{(1)}_{L_0}(u_i)}(\omega)$, $i=1,\ldots,n$. Then the  eigenvalues $E_{j_1\ldots j_n}$ of the non-interacting multi-particle random Hamiltonian\\ $\BH^{(n)}_{\BC^{(n)}_{L_0}(\Bu)}(\omega)$ are written as sums
\[
E_{j_1\ldots j_n}=\sum_{i=1}^n\lambda^{(i)}_{j_i}=\lambda^{(1)}_{j_1}+\cdots+\lambda^{(n)}_{j_n},
\]
while the corresponding eigenfunctions $\BPsi_{j_1\ldots j_n}$ can be chosen as tensor products
\[
\BPsi_{j_1\ldots j_n}=\psi^{(1)}_{j_1}\otimes\cdots\otimes\psi^ {(n)}_{j_n}.
\]
The eigenfunctions of finite volume Hamiltonians are assumed normalised.
\begin{theorem}\label{thm:initial.bound.np}
Let $1\leq n\leq N$ and $I_0\subset \DR$ a bounded interval. There exists $m^*>0$ such that for any cube  $\BC^{(n)}_{L_0}(\Bu)$ and all 
$E\in I_0$,
\begin{equation}\label{eq:initial.bound.np}
\prob{\text{$\BC^{(n)}_{L_0}(\Bu)$ is $(E,m^*,0)$-S}}
\le \frac{1}{2} L_0^{-2p^*4^{N-n}},
\end{equation}
with $L_0$ large enough and $p^*>6Nd$.
\end{theorem}

The proof of Theorem \ref{thm:initial.bound.np} relies on the following auxiliary statement.

\begin{lemma}\label{lem:1p.NS.implies.np.NS}
Let be given $N\geq n\geq 2$, $m^*>0$, a cube
$\BC^{(n)}_{L_0}(\Bu)$ and $E\in \DR$.
Suppose that $\BC^{(n)}_{L_0}(\Bu)$ is $E$-NR, and for any operator
$H^{(1)}_{C_{L_0}(u_i)}$,
all its eigenfunctions $\psi_{j_i}$ satisfy
\begin{equation}\label{eq:psi.m.loc}
\|\Bone_{C^{(1)}_{L_0}(u_i)}\cdot \psi_{j_i}\Bone_{C^{(1,int)}_{L_0}(u_i)}\| \le \ee^{-2\gamma(m^*,L_0,n)L_0}.
\end{equation}
Then $\BC^{(n)}_{L_0}(\Bu)$ is $(E,m^*,0)$-NS provided that $L_0 \ge L_*(m^*,N,d)$.
\end{lemma}

\begin{proof}

We choose the multi-particle eigenfunctions as tensor products of those of the single-particle Hamiltonian $H^{(1)}_{C^{(1)}_{L_0}(u_i)}$, $i=1,\ldots,n$, i.e., $\BPsi_j = \varphi^{(1)}_{j}\otimes\cdots\otimes \varphi^{(n)}$, corresponding to the eigenvalue,
$E_j = \lambda_j^{(1)}+\cdots+\lambda^{(n)}_j$.   
Now we have that 
\[
\begin{aligned} 
\BG^{(n)}_{\BC^{(n)}_{L_0}(\Bu)}(E)
&= \sum_{E_j\in\sigma(\BH^{(n)}_{\BC^{(n)}_{L_0}(\Bu)})} \BP_{\varphi^{(1)}_{j}}\otimes\cdots\otimes\BPsi\varphi_j^{(n-1)}G^{(1)}_{C^{(1)}_L(u_n)}(E-\lambda_{\neq n})\\
\end{aligned}
\]
where $\lambda_{\neq n}=\sum_{1\leq i\leq n-1} \lambda_i$ so that 

\begin{gather*}
\Bone_{\BC^{(n,out)}_{L_0}(\Bu)} \BG_{\BC^{(n)}_{L_0}(\Bu)}(E) \Bone_{\BC^{(n,int)}_{L_0}(\Bu)}\leq \Bone_{\BC^{(n)}_L(\Bu)}\BG^{(n)}_{\BC^{(n)}_L(\Bu)}(E)\Bone_{\BC^{(n)}_L(\Bu)}\\
\leq \sum_{i=1}^n \Bone^{\otimes^{(i-1)}}\otimes \Bone_{C^{(1)}_L(u_i)}\otimes\Bone^{\otimes(n-i)}\BG^{(n)}_{\BC^{(n)}_L(\Bu)}(E)\\
\leq \sum_{i=1}^n\left[\sum_j \Bone^{\otimes(i-1)}\otimes\Bone_{C^{(1)}_L(u_i)}\otimes\Bone^{\otimes(n-i)}\BP_{\varphi_j^{(1)}}\otimes\cdots\otimes\BPsi_{\varphi_j^{(n-1)}} G^{(1)}_{C^{(1)}_L(u_n)}(E-\lambda_{\neq n})\right].
\end{gather*}

By the Weyl's law, there exists $E^*>0$ which can be choosen arbitrarily large such that $\lambda_j\geq E^*$ for all $j\geq j^*=C_{Weyl}|C^{(1)}_{L_0}(u_n)|$.  Therefore, we divide the above sum on $j$ into two sums as follows:
\begin{gather*}
\Bone_{\BC^{(n,out)}_{L_0}(\Bu)} \BG_{\BC^{(n)}_{L_0}(\Bu)}(E) \Bone_{\BC^{(n,int)}_{L_0}(\Bu)}
\leq\sum_{i=1}^n\left(\sum_{j\leq j^*}+\sum_{j\geq j^*}\right)\\
\times \Bone^{\otimes(i-1)}\otimes\Bone_{C^{(1)}_L(u_i)}\otimes\Bone^{\otimes(n-i)}\BP_{\varphi_j^{(1)}}\otimes\cdots\otimes\BPsi_{\varphi_j^{(n-1)}} G^{(1)}_{C^{(1)}_L(u_n)}(E-\lambda_{\neq n}). 
\end{gather*}
Since 
\begin{gather*}
\|\Bone^{\otimes(i-1)}\otimes\Bone_{C^{(1)}_L(u_i)}\otimes\Bone^{\otimes(n-i)}\BP_{\varphi_j^{(1)}}\otimes\cdots\otimes\BPsi_{\varphi_j^{(n-1)}} G^{(1)}_{C^{(1)}_L(u_n)}(E-\lambda_{\neq n})\|\\
\leq \| \Bone_{C^{(1)}_L(u_i)}\cdot\varphi_j^{(1)}\|\cdot\ee^{L^{1/2}}
\leq \ee^{-2\gamma(m^*,L,1)L+L^{1/2}},
\end{gather*}
for $L>L^*(N,d,C_{Weyl})>0$ large enough and where we used the hypotheses on the exponential decay of the eigenfunctions of the single-paticle Hamiltonian. Thus, the infinite sum can made as small as an exponential decay provided that the length $L_0$ is large enough,
\[
\sum_{j\geq j^*} \|\Bone^{\otimes(i-1)}\otimes\Bone_{C^{(1)}_L(u_i)}\otimes\Bone^{\otimes(n-i)}\BP_{\varphi_j^{(1)}}\otimes\cdots\otimes\BPsi_{\varphi_j^{(n-1)}} G^{(1)}_{C^{(1)}_L(u_n)}(E-\lambda_{\neq n})\|\leq \frac{1}{2}\ee^{-\gamma(m^*,L_0,n)L_0}
\]
while the finite sum can be bounded by:
\[
n\cdot C_{Weyl}\cdot|C^{(1)}_L(u)|\cdot\ee^{-2\gamma(m^*,L,n)L}\ee^{L^1/2}\leq \frac{1}{2} \ee^{-\gamma(m^*,L,n)L},
\]
for $L_0>L^{**}$ with $L^{**}>0$ large enough. Finally, we obtain
\[
\|\Bone_{\BC^{(n,out)}_{L_0}(\Bu)} \BG_{\BC^{(n)}_{L_0}(\Bu)}(E) \Bone_{\BC^{(n,int)}_{L_0}(\Bu)}\|\leq \ee^{-\gamma(m^*,L,n)L},
\]
which proves the Lemma. 
\end{proof}

Now, we turn to the proof of Theorem \ref{thm:initial.bound.np}.
\begin{proof}[Proof of Theorem \ref{thm:initial.bound.np}]
Recall that by the single-particle Anderson localization theory, there exists $\widetilde{\mu}>0$ such that we have the following bounds on the exponential decay of the eigenfunctions: for all $u\in\DZ^d$,
\begin{equation}\label{eq:1p.loc}
\|1_{C^{(1)}_{L_0}(u)}\cdot \psi\|\leq \ee^{-\widetilde{\mu}|u|}.
\end{equation}
 Set $m^*=2^{-N-1}\cdot\widetilde{\mu}$ and introduce the events:
\begin{align*}
&\mathcal{N} := \{\exists i=1,\ldots,n: \exists \lambda_j\in\sigma(H^{(1)}_{C^{(1)}_{L_0}(u_i)}(\omega)): \text{ $\|1_{C^{(1)}_{L_0}(u_i)}\cdot\phi_j(u_i)\|>\ee^{-2\gamma(m^*,L_0,n)L_0}$} \},
\\
&\CR := \{ \BC^{(n)}_{L_0}(\Bu) \text{ is $E$-R }  \}.
\end{align*}
Then by Lemma \ref{lem:1p.NS.implies.np.NS}, Eqn. \eqref{eq:1p.loc} and theorem \ref{thm:Wegner} (A), we have:
\begin{align*}
\prob{\BC^{(n)}_{L_0}(\Bu) \text{ is $(E,m^*,0)$-S}}
&\le \prob{\mathcal{N}} + \prob{\CR},\\
&\leq  \sum_{i=1}^n \sum_{j\geq0}\frac{\esm{\|\Bone_{C^{(1)}_L(u_i)}\varphi_j\|}}{\ee^{-2\gamma(m^*,L,n)L}} +\prob{\CR}\\
&\leq\sum_{i=1}^n \sum_{j\geq0} \ee^{(-\widetilde{\mu}_1+2\gamma(m^*,L_0,n))L_0}+ L_0^{-4^N\,p}.
\end{align*}
Since  $2\gamma(m^*,L_0,n)< 2^{N+1}m^*=\widetilde{\mu}_1$, $\widetilde{\mu_1}-2\gamma(m^*,L,n)>0$. Using the Weyl's law, we can divide the infinite sum above into two sums. Namely, there exists $E^*>0$ arbitrarily large such that $\lambda^{(1)}_j\geq E^*$ for $j\geq j^*=|C^{(1)}_{L_0}(u_i)|$ which yields
\[
 \sum_{j\geq0} \ee^{(-\widetilde{\mu}_1+2\gamma(m^*,L_0,n))L_0}=\left(\sum_{j\leq j^*}+\sum_{j>j^*}\right)\ee^{(-\widetilde{\mu}_1+2\gamma(m^*,L_0,n))L_0}.
\]
Above, the infinite sum can be made small than any polynomial power law provided that $L_0$ is large enough. We have
\[
\sum_i\left(\sum_{j\leq j^*}+\sum_{j>j^*}\right)\ee^{(-\widetilde{\mu}_1+2\gamma(m^*,L_0,n))L_0}\leq \frac{1}{3}L_0^{-2p4^{N-n}}+\frac{1}{3}L_0^{-2p4^{N-n}} +L_0^{-p4^N}<L_0^{-2p4^{N-n}}.
\] 	 
\end{proof}
We state and give here the proof of some important results from the paper \cite{E13} which use the fact that we are in the weakly interacting regime.
The constant $m^*>0$ is the one from Theorem \ref{thm:initial.bound.np}.
\subsection{The fixed energy  MSA bound for weakly interacting multi-particle systems}\label{sec:weak.int.initial.bound}
Now we derive the required initial estimate from its counterpart established for non-interacting  systems.

\begin{theorem}\label{thm:weak.interaction.fixed}
Let $1\leq n\leq N$. Suppose that the Hamiltonians $\BH_{0}^{(n)}(\omega)$ (without inter-particle interaction) fulfills the following condition: for all $E\in I$ and all $\Bu\in\DZ^{nd}$
\begin{equation}\label{eq:lem.weak.U.1}
\prob{\BC_{L_0}^{(n)}(\Bu)\text{ is $(E,m^*,0)$-S}}\leq\frac{1}{2} L_0^{-2p^*4^{N-n}},\qquad\text{with } 
p^*>6Nd.
\end{equation}
Then there exists $h^*>0$ such that for all $h\in(-h^*, h^*)$ the Hamiltonian $\BH_{h}^{(n)}(\omega)$, with interaction of amplitude $|h|$, satisfies a similar bound: there exist some $p>6Nd, m>0$ such that for all $E\in I$ and all $\Bu\in\DZ^{nd}$
\[
\prob{\BC_{L_0}^{(n)}(\Bu)\text{ is $(E,m,h)$-S}}\leq\frac{1}{2} L_0^{-2p\,4^{N-n}}.
\]
\end{theorem}

\begin{proof} First observe that the result of \eqref{eq:lem.weak.U.1} is proved in the statemeent of Theorem \ref{thm:initial.bound.np}.
Set
\[
\BG_{\BC_{L_0}^{(n)}(\Bu),h}(E)=(\BH^{(n)}_{\BC_{L_0}^{(n)}(\Bu),h}-E)^{-1}, \; h\in\DR.
\]
By definition, a cube $\BC_{L_0}^{(n)}(\Bu)$ is $(E,m^*,0)$-NS iff
\begin{equation}\label{eq:proof.lem.weak.U.0}
\|\Bone^{(n,out)}_{\Bu}\BG^{(n)}_{\BC^{(n)}_L(\Bu)}(E)\Bone^{(n,int)}_{\Bu}\|\leq\ee^{-\gamma(m,L_0,n)L_0},
\end{equation}

Therefore, there exists sufficiently small $\epsilon>0$ such that
\begin{equation}\label{eq:proof.lem.weak.U.1}
\|\Bone^{(n,out)}_{\Bu}\BG^{(n)}_{\BC^{(n)}_L(\Bu)}(E)\Bone^{(n,int)}_{\Bu}\|\leq\ee^{-\gamma(m,L,n)L}-\epsilon,
\end{equation}
where $m= m^*/2>0$. Since, by assumption, $p^*>6Nd$, there exists $6Nd<p<p^*$ and $\tau>0$ such that $L_0^{-2p4^{N-n}}-\tau>L_0^{-2p^*4^{N-n}}$. With such values $p$ and $\tau$, inequality \eqref{eq:lem.weak.U.1} with $p^*>6Nd$ implies
\begin{equation}\label{eq:lem.weak.U.p.prime}
\DP\{\BC_{L_0}^{(n)}(\Bu)\text{ is $(E,m^*,0)$-S}\}<\frac{1}{2}L_0^{-2p4^{N-n}}-\frac{1}{2}\tau.
\end{equation}
Next, it follows from the second resolvent identity that
\begin{equation}\label{eq:proof.lem.weak.U.2}
\|\BG_{\BC_{L_0}^{(n)}(\Bu),0}(E) - \BG_{\BC_{L_0}^{(n)}(\Bu),h}(E) \|
\leq |h|\, \|\BU\| \cdot \|\BG_{\BC_{L_0}^{(n)}(\Bu),0}(E)\| \cdot \|\BG_{\BC_{L_0}^{(n)}(\Bu),h}(E)\|.
\end{equation}
By Theorem \ref{thm:Wegner}, applied to Hamiltonians
$\BH^{(n)}_{\BC_{L_0}^{(n)}(\Bu),0}$ and $\BH_{\BC_{L_0}^{(n)}(\Bu),h}^{(n)}$, for any $\tau>0$ there is $B(\tau)\in(0,+\infty)$ such that
\begin{align*}
&\DP\bigl\{\|\BG_{\BC_{L_0}^{(n)}(\Bu),0}(E)\|\geq B(\tau)\bigr\}\leq\frac{\tau}{4}\,,\\
&\DP\bigl\{\|\BG_{\BC_{L_0}^{(n)}(\Bu),h}(E)\|\geq B(\tau)\bigr\}\leq\frac{\tau}{4}\,.
\end{align*}
Therefore,
\begin{align*}
&\DP\bigl\{\|\BG_{\BC_{L_0}^{(n)}(\Bu),0}(E)-\BG_{\BC_{L_0}^{(n)}(\Bu),h}(E)\|
\geq |h|\,\|\BU\|B^{2}(\tau)\bigr\}\\
&\qquad\leq\DP\bigl\{\|\BG_{\BC_{L_0}^{(n)}(\Bu),0}(E)\|\geq B(\tau)\bigr\}+\DP\bigl\{\|\BG_{\BC_{L_0}^{(n)}(\Bu),h}(E)\|\geq B(\tau)\bigr\}\\
&\qquad\leq 2\,\frac{\tau}{4}=\frac{\tau}{2}\,.
\end{align*}
Set $h^*:=\frac{\epsilon}{2\|\BU\|(B(\tau))^2}>0$. We see that if $|h|\leq h^*$, then $|h|\times\|\BU\|\times(B(\tau))^2\leq\frac{\epsilon}{2}\,$.
Hence,
\begin{equation}\label{eq.proof.lem.weak.U.3}
\DP\bigl\{\|\BG_{\BC_{L_0}^{(n)}(\Bu),0}-\BG_{\BC_{L_0}^{(n)}(\Bu),h}\|\geq\frac{\epsilon}{2}\bigr\}\leq 2\,\frac{\tau}{4}\,.
\end{equation}
Combining \eqref{eq:proof.lem.weak.U.1}, \eqref{eq:lem.weak.U.p.prime}, and
\eqref{eq.proof.lem.weak.U.3}, we obtain that for all $E\in I$
\begin{align*}
&\DP\bigl\{ \BC_{L_0}^{(n)}(\Bu) \text{ is $(E,m,h)$-S}\bigr\}\\
&\quad\leq\DP\bigl\{\BC_{L_0}^{(n)}(\Bu)\text{ is $(E,m^*,0)$-S}\bigr\}\\
&+\DP\bigl\{\|\BG_{\BC_{L_0}^{(n)}(\Bu),0}(E)-\BG_{\BC_{L_0}^{(n)}(\Bu),h}(E)\|\geq\frac{\epsilon}{2}\bigr\}\\
&\quad\leq\bigl(\frac{1}{2}L_0^{-2p4^{N-n}}-\frac{1}{2}\tau\bigr)+\frac{\tau}{2}=\frac{1}{2}L_0^{-2p'4^{N-n}}.\qedhere
\end{align*}
\end{proof}

\subsection{The variable energy MSA bound for weakly interacting multi-particle systems}
Here, we deduce from the fixed energy bound, the variable energy initial multi-scale analysis bound for the weakly interacting multi-particle system. 
We will prove localization in each compact interval $I_0$ of the following form: let $E_0\in \DR$ and $\delta:=\frac{1}{2}\ee^{-2L_0^{1/2}}(\ee^{-m_1L_0}-\ee^{-mL_0})$ where $0<m_1<m$ by definition. Set
\[
I_0:=[E_0-\delta;E_0+\delta].
\]
The result on the variable energy MSA is given below in
\begin{theorem}\label{thm:initial.var.energy}
Let $1\leq n\leq N$. For any $\Bu\in\DZ^{nd}$ we have
\begin{equation}
\prob{\exists E\in I_0: \text{$\BC^{(n)}_{L_0}(\Bu)$ is $(E,m_1)$-S}}\leq L_0^{-2p\,4^{N-n}},
\end{equation}
for some $m_1>0$.
\end{theorem}

\begin{proof}
Let $E_0\in I$. By the resolvent equation
\[
\BG_{\BC^{(n)}_{L_0}(\Bu),h}(E)=\BG_{\BC^{(n)}_{L_0}(\Bu),h}(E_0)+ (E-E_0)\BG_{\BC^{(n)}_{L_0}(\Bu),h}(E)\BG_{\BC^{(n)}_{L_0}(\Bu),h}(E_0).
\]
If $\dist(E_0,\sigma(\BH^{(n)}_{\BC^{(n)}_{L_0}(\Bu),h}))\geq \ee^{-L_0^{1/2}}$ and $|E-E_0|\leq \frac{1}{2} \ee^{-L_0^{1/2}}$, then $\dist(E,\sigma(\BH^{(n)}_{\BC^{(n)}_{L_0}(\Bu),h}))\geq \frac{1}{2} \ee^{-L_0^{1/2}}$.

If in addition, $\BC^{(n)}_{L_0}(\Bu)$ is $(E_0,m,h)$-NS, then 
\[
\|\Bone^{(n,out)}_{\Bx}\BG^{(n)}_{\BC^{(n)}_L(\Bx)}(E)\Bone^{(n,int)}_{\Bx}\|\leq \ee^{-m(1+L_0^{-1/8})^{N-n+1}L_0} + 2 |E-E_0| \ee^{2L_0^{1/2}}.
\]
Therefore, for $m_1=\frac{m}{2}$, if we put 
\[
\delta=\frac{1}{2}\ee^{-2L_0^{1/2}}(\ee^{-m_1(1+L_0^{-1/8})^{N-n+1}L_0}- \ee^{-m(1+L_0^{-1/8})^{N-n+1}L_0}),\quad I_{0}=[E_0-\delta,E_0+\delta],
\]
we have that 
\begin{align*}
&\prob{\text{$\exists E\in I_0$, $\BC^{(n)}_{L_0}(\Bu)$ is $(E,m_1,h)$-S}}\leq \prob{\text{$\BC^{(n)}_{L_0}(\Bu)$ is $(E_0,m,h)$-S}}\\
&\qquad + \prob{\dist(E_0,\sigma(\BH^{(n)}_{\BC^{(n)}_{L_0}(\Bu)}))\leq \ee^{-L_0^{1/2}}}\\
&\leq \frac{1}{2} L_0^{-2p4^{N-n}} + L_0^{-p4^{N}}<L_0^{-2p4^{N-n}}.
\end{align*}
We used Theorem \ref{thm:weak.interaction.fixed} to bound the first term and the Wegner estimate Theorem \ref{thm:Wegner} (A) to bound the other term.
\end{proof}

Below, we develop the indcution step of the multi-scale analysis and for the reader convenience we also give the proof of some important results.

\section{Multi-scale induction}

In the rest of the paper, we assume that  $n\geq 2$ and  $I_0$ is the interval from the previous section. 

Recall the following facts from \cite{E12}: Consider a cube $\BC^{(n)}_{L}(\Bu)$, with $\Bu=(u_1,\ldots,u_n)\in(\DZ^d)^n$.  We define
\[
\varPi\Bu=\{u_1,\ldots,u_n\},
\]
and 
\[
\varPi\BC^{(n)}_L(\Bu)=C^{(1)}_L(u_1)\cup\cdots\cup C^{(1)}_L(u_n).
\]
\begin{definition}
Let $L_0>3$ be a constant and $\alpha=3/2$. We define the sequence $\{L_k: k\geq 1\}$ recursively  as follows:
\[
L_k:=\lfloor L_{k-1}^{\alpha}\rfloor +1, \qquad \text{for all $k\geq 1$}.
\]
\end{definition}

Let $m>0$ a positive constant, we also introduce the following property, namely the multi-scale analysis bounds at any scale length $L_k$, and for any pair of separable cubes $\BC^{(n)}_{L_k}(\Bu)$ and $\BC^{(n)}_{L_k}(\Bv)$,
\begin{dsknn*}
\[
\prob{\exists E\in I_0: \text{$\BC^{(n)}_{L_k}(\Bu)$ and $\BC^{(n)}_{L_k}(\Bv)$ are $(E,m)$-S}}\leq L_k^{-2p4^{N-n}},
\]
where $p>6Nd$.
\end{dsknn*}

In both the single-particle and the multi-particle system, given the results on the multi-scale analysis property $\dsknn$ above one can deduce the localization results see for example the papers \cites{DK89,DS01} for those concerning the single-particle case and \cites{E12,CS09a} for multi-particle systems. We have the following

\begin{theorem}\label{thm:EFC.bound}
For any $1\leq n'<n$, assume that property $\dsk{n',N}$ holds true for all $k\geq 0$, then there exists a constant $\widetilde{\mu}>0$ such that for any cube $\BC^{(n')}_{L}(\Bu')$
\begin{equation}\label{eq:EFC.bound}
\esm{ \|\Bone_{\BC^{(n',out)}_{L}(\Bu')}\BG^{(n')}_{\BC^{(n')}_L(\Bu')}(E)\Bone_{\BC^{(n',int)}_L(\Bu)}\|} \leq\ee^{-\widetilde{\mu}L}.
\end{equation}
\end{theorem}

\begin{definition}[fully/partially interactive]\label{def:diagonal.cubes}
An $n$-particle cube $\BC_L^{(n)}(\Bu)\subset\DZ^{nd}$ is
called fully interactive (FI) if
\begin{equation}\label{eq:def.FI}
\diam \varPi \Bu := \max_{i\ne j} |u_i - u_j| \le n(2L+r_0),
\end{equation}
and partially interactive (PI) otherwise. 
\end{definition}

The following simple statement clarifies the notion of PI cube.

\begin{lemma}\label{lem:PI.cubes}
If a cube $\BC_L^{(n)}(\Bu)$ is PI, then there exists a subset $\CJ\subset\left\{1,\dots,n\right\}$ with $1\leq\card\CJ\leq n-1$ such that
\[
\dist\left(\varPi_{\CJ}\BC_L^{(n)}(\Bu),\varPi_{\CJ^{\comp}}\BC_L^{(n)}(\Bu)\right)>r_0,
\]
\end{lemma}  

\begin{proof}
See the appendix section \ref{sec:appendix}.
\end{proof}

If $\BC^{(n)}_L(\Bu)$ is a PI cube by the above Lemma, we can write it as 
\begin{equation}\label{eq:cartesian.cubes}
\BC_L^{(n)}(\Bu)=\BC_L^{(n')}(\Bu')\times\BC_L^{(n'')}(\Bu''),
\end{equation}
with 
\begin{equation}\label{eq:cartesian.cubes.2}
\dist\left(\varPi\BC_L^{(n')}(\Bu'),\varPi\BC_L^{(n'')}(\Bu'')\right)>r_0,
\end{equation}
where $\Bu'=\Bu_{\CJ}=(u_j:j\in\CJ)$, $\Bu''=\Bu_{\CJ^{\comp}}=(u_j:j\in\CJ^{\comp})$, $n'=\card \CJ$ and $n''=\card \CJ^{\comp}$. 

Throughout, when we write a PI cube $\BC^{(n)}_L(\Bu)$ in the form \eqref{eq:cartesian.cubes}, we implicitly assume that the projections satisfy \eqref{eq:cartesian.cubes.2}.
Let $\BC^{(n')}_{L_k}(\Bu')\times\BC^{(n'')}_{L_k}(\Bu'')$ be the decomposition of the PI cube $\BC^{(n)}_{L_k}(\Bu)$ and $\{\lambda_i,\varphi_i\}$ and $\{\mu_j,\phi_j\}$ be the eigenvalues and corresponding eigenfunctions of $\BH_{\BC_{L_k}^{(n')}(\Bu')}^{(n')}$ and $\BH_{\BC_{L_k}^{(n'')}(\Bu'')}^{(n'')}$ respectively. Next, we  can choose the eigenfunctions $\BPsi_{ij}$ of $\BH_{\BC^{(n)}_{L_k}(\Bu)}(\omega)$ as tensor products:
\[
\BPsi_{ij}=\varphi_i\otimes\phi_j
\]
The eigenfunctions appearing in subsequent arguments and calculation will be assumed normalized.

Now we turn to geometrical properties of FI cubes.
\begin{lemma}\label{lem:FI.cubes}
Let $n\geq 1$, $L>2r_0$ and consider two FI cubes $\BC_L^{(n)}(\Bx)$ and $\BC_L^{(n)}(\By)$ with $|\Bx-\By|>7\,nL$. Then
\begin{equation}
\varPi\BC_L^{(n)}(\Bx)\cap\varPi\BC_L^{(n)}(\By)=\varnothing.
\end{equation}
\end{lemma}

\begin{proof}
See the appendix section \ref{sec:appendix}.
\end{proof}

Given an $n$-particle cube $\BC_{L_{k+1}}^{(n)}(\Bu)$ and $E\in\DR$, we denote
\begin{itemize}
\item
by $M_{\pai}^{\sep}(\BC_{L_{k+1}}^{(n)}(\Bu),E)$ the maximal number of pairwise separable, 
$(E,m)$-singular PI cubes $\BC_{L_k}^{(n)}(\Bu^{(j)})\subset\BC_{L_{k+1}}^{(n)}(\Bu)$;
\item
by  $M_{\pai}(\BC_{L_{k+1}}^{(n)}(\Bu),E)$ the maximal number of (not necessarily separable)
$(E,m)$-singular PI cubes $\BC^{(n)}_{L_k}(\Bu^{(j)})$ contain in $\BC^{(n)}_{L_{k+1}}(\Bu)$ with $\Bu^{(j)}, \Bu^{(j')}\in\DZ^{nd}$ and $|\Bu^{(j)}-\Bu^{(j')}|>7NL_k$ for all $j\neq j'$;
\item
by $M_{\fui}(\BC_{L_{k+1}}^{(n)}(\Bu),E)$ the maximal number of 
$(E,m)$-singular FI cubes  $\BC_{L_k}^{(n)}(\Bu^{(j)})\subset \BC_{L_{k+1}}^{(n)}(\Bu)$ with $|\Bu^{(j)}-\Bu^{(j')}|>7NL_k$ for all $j\neq j'$\footnote{Note that by lemma \ref{lem:FI.cubes}, two FI cubes $\BC^{(n)}_{L_k}(\Bu^{(j)})$ and $\BC^{(n)}_{L_k}(\Bu^{(j')})$ with $|\Bu^{(j)}-\Bu^{(j')}|>7NL_k $ are automatically separable.},
\item
$M_{\pai}(\BC_{L_{k+1}}^{(n)}(\Bu),I):=\sup_{E\in I}M_{\pai}(\BC_{L_{k+1}}^{(n)}(\Bu),E)$.
\item
$M_{\fui}(\BC_{L_{k+1}}^{(n)}(\Bu),I):=\sup_{E\in I}M_{\fui}(\BC_{L_{k+1}}^{(n)}(\Bu),E)$.
\item
by $M(\BC_{L_{k+1}}^{(n)}(\Bu),E)$ the maximal number of 
$(E,m)$-singular cubes  $\BC_{L_k}^{(n)}(\Bu^{(j)})\subset \BC_{L_{k+1}}^{(n)}(\Bu)$ with $\dist(\Bu^{(j)},\partial\BC^{(n)}_{L_{k+1}}(\Bu))\geq 2L_k$ and $|\Bu^{(j)}-\Bu^{(j')}|>7NL_k$ for all $j\neq j'$.
\item
by $M^{\sep}(\BC_{L_{k+1}}^{(n)}(\Bu),E)$ the maximal number of pairwise separable
$(E,m)$-singular cubes  $\BC_{L_k}^{(n)}(\Bu^{(j)})\subset \BC_{L_{k+1}}^{(n)}(\Bu)$ 
\end{itemize}
Clearly
\[
M_{\pai}(\BC_{L_{k+1}}^{(n)}(\Bu),E)+M_{\fui}(\BC_{L_{k+1}}^{(n)}(\Bu),E)\geq M(\BC_{L_{k+1}}^{(n)}(\Bu),E).
\]

\subsection{Pairs of partially interactive cubes}\label{ssec:PI.cubes}
		
		Let  $\BC_{L_{k+1}}^{(n)}(\Bu)=\BC^{(n')}_{L_{k+1}}(\Bu')\times\BC^{(n'')}_{L_{k+1}}(\Bu'')$ be a PI-cube. We also write $\Bx=(\Bx',\Bx'')$ for any point $\Bx\in\BC_{L_{k+1}}^{(n)}(\Bu)$, in the same way as $\Bu=(\Bu',\Bu'')$. So  the corresponding Hamiltonian $\BH^{(n)}_{\BC_{L_{k+1}}^{(n)}(\Bu)}$ is written in the form:
\begin{equation}\label{eq:decomp,H}
\BH_{\BC_{L_{k+1}}^{(n)}(\Bu)}^{(n)}\BPsi(\Bx)=(-\BDelta\BPsi)(\Bx)+\left[\BU(\Bx')+\BV(\Bx',\omega)+\BU(\Bx'')+\BV(\Bx'',\omega)\right]\BPsi(\Bx)
\end{equation}
or, in compact form 
\[
\BH_{\BC_{L_{k+1}}^{(n)}(\Bu)}^{(n)}=\BH_{\BC_{L_{k+1}}^{(n')}(\Bu')}^{(n')}\otimes\mathbf{I}+ \mathbf{I}\otimes \BH_{\BC_{L_{k+1}}^{(n'')}(\Bu'')}^{(n'')}.
\]

\begin{definition}\label{def:localized}
Let $n\geq 2$ and  $\BC^{(n')}_{L_k}(\Bu')\times\BC^{(n'')}_{L_k}(\Bu'')$ be the decomposition of the PI cube $\BC^{(n)}_{L_k}(\Bu)$. Then  $\BC^{(n)}_{L_k}(\Bu)$ is called
\begin{enumerate}[(i)]
\item
$m$-left-localized if for  any normalized eigenfunction $\varphi^{(n')}$ of the restricted hamiltonian $\BH^{(n')}_{\BC^{(n')}_{L_k}(\Bu')}(\omega)$, we have
\[
\|\Bone_{\BC^{(n',out)}_L(\Bu')}\varphi\|\leq \ee^{-2\gamma(m,L_k,n')L_k},
\]
otherwise, it is called $m$-non-left-localized,
\item
$m$-right-localized if for any normalized eigenfunction $\phi^{(n'')}$ of the restricted hamiltonian $\BH^{(n'')}_{\BC^{(n'')}_{L_k}(\Bu'')}(\omega)$, we have
\[
\|\Bone_{\BC^{(n'')}_L(\Bu'')}\varphi^{(n'')}\|\leq \ee^{-2\gamma(m,L_k,n'')L_k},
\]
otherwise, it is called $m$-non-right-localized,
\item
$m$-localized if  it is $m$-left-localized and $m$-right-localized.
Otherwise it is called $m$-non-localized. 
\end{enumerate}
\end{definition}

\begin{lemma} \label{lem:localized}
Let $E\in I$ and $\BC_{L_k}^{(n)}(\Bu)$ be a PI cube. Assume that
$\BC_{L_k}^{(n)}(\Bu)$ is $E$-NR and $m$-localized.
Then $\BC^{(n)}_{L_k}(\Bu)$ is $(E,m)$-NS.
\end{lemma}

\begin{proof}
We proceed as in Lemma \ref{lem:1p.NS.implies.np.NS}.
\end{proof}

Now, before proving the main result of this subsection concerning the probability of two PI cubes to be singular at the same energy, we need first to estimate the one of a non localized cube given in the statement below. 

\begin{lemma}\label{lem:prob.localized}
Let $\BC^{(n)}_{L_k}(\Bu)$ be a PI cube. Then
\[
\prob{ \BC^{(n)}_{L_k}(\Bu) \text{ is $m$-non localized}}\leq \frac{1}{2}L_k^{-4p\,4^{N-n}}.
\]
\end{lemma}
\begin{proof}
The proof combines the ideas of Theorem \ref{thm:initial.bound.np} in the case of the multi-particle system without interaction and the induction assertion on localization given in Theorem \ref{thm:EFC.bound}.
\end{proof}

Now, we state the main result of this subsection, i.e., the probability bound of two PI cubes to be singular at the same energy belonging to the  compact interval $I_0$ introduced at the beginning of the section. 

\begin{theorem}\label{thm:partially.interactive}
Let $2\leq n\leq N$. There exists $L_1^*=L_1^*(N,d)>0$ such that if $L_0\geq L_1^*$ and if for $k\geq 0$ $\dsn{k,n'}$ holds true for any $1\leq n'<n$, then $\dskonn$ holds true
for any pair of separable PI cubes $\BC_{L_{k+1}}^{(n)}(\Bx)$ and $\BC_{L_{k+1}}^{(n)}(\By)$.
\end{theorem}

\begin{proof}
Let $\BC_{L_{k+1}}^{(n)}(\Bx)$ and $\BC_{L_{k+1}}^{(n)}(\By)$ be two separable PI-cubes. Consider the events:
\begin{align*}
\rB_{k+1}
&=\bigl\{\exists\,E\in I_0:\BC_{L_{k+1}}^{(n)}(\Bx)\text{ and $\BC_{L_{k+1}}^{(n)}(\By)$ are $(E,m)$-S}\bigr\},\\
\rR
&=\bigl\{\exists\,E\in I_0:\text{ $\BC_{L_{k+1}}^{(n)}(\Bx)$ and $\BC_{L_{k+1}}^{(n)}(\By)$ are $E$-R}\bigr\},\\
\mathcal{N}_{\Bx}
&=\bigl\{ \BC_{L_{k+1}}^{(n)}(\Bx)\text{ is $m$-non-localized}\bigr\},\\
\mathcal{N}_{\By}
&=\bigl\{\BC_{L_{k+1}}^{(n)}(\By)\text{ is $m$-non-localized}\bigr\}.
\end{align*}
If $\omega\in\rB_{k+1}\setminus\rR$, then $\forall E\in I_0$, $\BC_{L_{k+1}}^{(n)}(\Bx)$ or $\BC_{L_{k+1}}^{(n)}(\By)$ is $E$-NR. If $\BC_{L_{k+1}}^{(n)}(\By)$ is $E$-NR, then it must be $m$-non-localized: otherwise it would have been $(E,m)$-NS by Lemma~\ref{lem:localized}. Similarly, if $\BC_{L_{k+1}}^{(n)}(\Bx)$ is $E$-NR, then it must be $m$-non-localized. This implies that
\[
\rB_{k+1}\subset\rR\cup\mathcal{N}_{\Bx}\cup\mathcal{N}_{\By}.
\]
Therefore, using Theorem \ref{thm:Wegner} and Lemma \ref{lem:prob.localized}, we have
\begin{align*}
\DP\left\{\rB_{k+1}\right\}&\leq\DP\left\{\rR\right\}+\DP\{\mathcal{N}_{\Bx}\}+\DP\{\mathcal{N}_{\By}\}\\
&\leq L_{k+1}^{-p4^{N}}+\frac{1}{2}L_{k+1}^{-4p\,4^{N-n}}+\frac{1}{2}L_{k+1}^{-4p\,4^{N-n}}.\\
\end{align*}
 Finally 
\begin{equation}\label{eq:bound.PI}
\prob{\rB_{k+1}}\leq L_{k+1}^{-4^N\,p}+L_{k+1}^{-4p4^{N-n}}<L_{k+1}^{-2p4^{N-n}},
\end{equation}
which proves the result.
\end{proof}

For subsequent calculations and proofs we give the following two Lemmas. 

\begin{lemma}\label{lem:MPI}
If $M(\BC^{(n)}_{L_{k+1}}(\Bu),E)\geq \kappa(n)+2$ with $\kappa(n)=n^n$, then $M^{\sep}(\BC^{(n)}_{L_{k+1}}(\Bu),E)\geq 2$.\\
\noindent Similarly, if $M_{\pai}(\BC^{(n)}_{L_{k+1}}(\Bu),E)\geq \kappa(n)+2$  then $M_{\pai}^{\sep}(\BC^{(n)}_{L_{k+1}}(\Bu),E)\geq 2$.
\end{lemma}
\begin{proof}
See the appendix section \ref{sec:appendix}.
\end{proof}

\begin{lemma}\label{lem:MND}
With the above notations, assume that $\dsn{k-1,n'}$ holds true for all $1\leq n'<n$ then
\begin{equation}\label{eq:MND}
\DP\left\{M_{\pai}(\BC_{L_{k+1}}^{(n)}(\Bu),I)\geq \kappa(n)+ 2\right\}\leq \frac{3^{2nd}}{2} L_{k+1}^{2nd}\left(L_k^{-4^Np}+L_k^{-4p\,4^{N-n}}\right).
\end{equation}
\end{lemma}	
\begin{proof}
See the appendix section \ref{sec:appendix}.
\end{proof}

\subsection{Pairs of fully interactive cubes}\label{ssec:FI.cubes}
Our aim now is to prove $\dskonn$ for a pair of separable fully interactive cubes $\BC_{L_{k+1}}^{(n)}(\Bx)$ and $\BC_{L_{k+1}}^{(n)}(\By)$. We adapt to the continuum a very crucial and hard result obtained in the paper \cite{E12} and which generalized to multi-particle systems some previous work by von Dreifus and Klein \cite{DK89} on the lattice and Stollmann \cite{St01} in the continuum for single-particle models. 

\begin{lemma}\label{lem:CNR.NS}
Let $J=\kappa(n)+5$ with $\kappa(n)=n^n$ and $E\in\DR$. Suppose that
\begin{enumerate}[\rm(i)]
\item
$\BC_{L_{k+1}}^{(n)}(\Bx)$ is $E$-CNR,
\item
$M(\BC_{L_{k+1}}^{(n)}(\Bx),E)\leq J$.
\end{enumerate}
Then there exists $\tilde{L}_2^*(J,N,d)>0$ such that if $L_0\geq \tilde{L}_{2}^*(J,N,d)$ we have that $\BC^{(n)}_{L_{k+1}}(\Bx)$ is $(E,m)$-NS.
\end{lemma}
\begin{proof}
Since, $M(\BC^{(n)}_{L_{k+1}}(\Bx);E)\leq J$, there exists at most $J$ cubes of side length $2L_k$ contained in $\BC^{(n)}_{L_{k+1}}(\Bx)$ that are $(E,m)$-S with centers at distance $> 7NL_k$. Therefore, we can find $\Bx_i\in\BC^{(n)}_{L_{k+1}}(\Bu)\cap\Gamma_{\Bx}$ with $\Gamma_{\Bx}=\Bx+\frac{L_k}{3}\DZ^{nd}$
\[
\dist(\Bx_i,\partial\BC^{(n)}_{L_{k+1}}(\Bx))\geq 2L_k,\quad\text{ $i=1,\ldots,r\leq J$},
\]
such that, if $\Bx_0\in \BC^{(n)}_{L_{k+1}}(\Bx)\setminus\bigcup_{i=1}^r\BC^{(n)}_{2L_k}(\Bx_i)$, then the cube $\BC^{(n)}_{L_k}(\Bx_0)$ is $(E,m)$-NS.

We do an induction procedure in $\BC^{(n,int)}_{L_{k+1}}(\Bx)$ and start with $\Bx_0\in\BC^{(n,int)}_{L_{k+1}}(\Bx)$. We estimate $\|\Bone_{\BC^{(n,out)}_{L_{k+1}}(\Bx)}\BG^{(n)}_{L_{k+1}}(E)\Bone_{\BC^{(n,int)}_{L_k}(\Bx_0)}\|$. Suppose that $\Bx_0,\ldots,\Bx_{\ell}$ have been choosen for $\ell\geq 0$. We have two cases:

\begin{enumerate}
\item[case(a)] $\BC^{(n)}_{L_k}(\Bx_{\ell})$ is $(E,m)$-NS.\\
 In this case, we apply the (GRI) Theorem \ref{thm:GRI.GF} and obtain
\begin{gather*}
\|\Bone_{\BC^{(n,out)}_{L_{k+1}}(\Bx)}\BG^{(n)}_{\BC^{(n)}_{L_{k+1}}(\Bx)}(E)\Bone_{\BC^{(n,int)}_{L_{k}}(\Bx_0)}\|\\
\leq C_{geom}\|\Bone_{\BC^{(n,out)}_{L_{k+1}}(\Bx)}\BG^{(n)}_{\BC^{(n)}_{L_{k+1}}(\Bx)}(E)\Bone_{\BC^{(n,out)}_{L_k}(\Bx_0)}\|\cdot\|\Bone_{\BC^{(n,out)}_{L_k}(\Bx)}\BG^{(n)}_{\BC^{(n)}_{L_k}(\Bx_0)}(E)\Bone_{\BC^{(n,int)}_{L_k}(\Bx_0)}\|\\
\leq C_{geom}\|\Bone_{\BC^{(n,out)}_{L_{k+1}}(\Bx)}\BG^{(n)}_{\BC^{(n)}_{L_{k+1}}}(E)\Bone_{\BC^{(n,out)}_{L_k}(\Bx)}\|\cdot\ee^{-\gamma(m,L_k,n)L_k}.
\end{gather*}
 We replace in the above analysis $\Bx$ with $\Bx_{\ell}$ and we get 
\[
\|\Bone_{\BC^{(n,out)}_{L_{k+1}}(\Bx_{\ell})}\BG^{(n)}_{\BC^{(n)}_{L_{k+1}}(\Bx_{\ell})}(E)\Bone_{\BC^{(n,out)}_{L_k}(\Bx_{\ell})}\|\leq 3^{nd}\|\Bone_{\BC^{(n,out)}_{L_{k+1}}(\Bx_{\ell})}\BG^{(n)}_{\BC^{(n)}_{L_{k+1}}\Bx_{\ell}}(E)\Bone_{\BC^{(n,int)}_{L_{k}}(\Bx_{\ell+1}}\|,
\]
where $\Bx_{\ell+1}$ is choosen in such a way that the norm in the right hand side in the above equation is maximal. Observe that $|\Bx_{\ell}-\Bx_{\ell+1}|=L_k/3$. We therefore obtain

\begin{gather*}
\|\Bone_{\BC^{(n,out)}_{L_{k+1}}(\Bx)}\BG^{(n)}_{\BC^{(n)}_{L_{k+1}}}(E)\Bone_{\BC^{(n,int)}_{L_k}(\Bx_{\ell}}\|\\
\leq C_{geom}3^{nd}\ee^{-\gamma(m,L_k,n)L_k}\cdot\|\Bone_{\BC^{(n,out)}_{L_{k+1}}(\Bx)}\BG^{(n)}_{\BC^{(n)}_{L_{k+1}}\Bx)}(E)\Bone_{\BC^{(n,int)}_{L_k}(\Bx_{\ell+1})}\|\\
\leq \delta_{+}\|\Bone_{\BC^{n,out)}_{L_{k+1}}(\Bx)}\BG^{(n)}_{\BC^{(n)}_{L_{k+1}}(\Bx)}(E)\Bone_{\BC^{(n,int)}_{L_k}(\Bx_{\ell+1})}\|
\end{gather*}
with 
\[
\delta_{+}=3^{nd}C_{geom}\ee^{-\gamma(m,L_k,n)L_k}.
\]

\item[case(b)] $\BC^{(n)}_{L_k}(\Bx_{\ell})$ is  $(E,m)$-S. \\
Thus, there exists $i_0=1,\ldots,r$ such that $\BC^{(n)}_{L_k}(\Bx_{\ell})\subset\BC^{(n)}_{2L_k}(\Bx_{i_0})$. We apply again the (GRI) this time with $\BC^{n)}_{L_{k+1}}(\Bx)$ and $\BC^{(n)}_{2L_k}(\Bx_{i_0})$ and obtain
\begin{gather*}
\|
\Bone_{\BC^{(n,out)}_{L_{k+1}}(\Bx)}\BG^{(n)}_{\BC^{(n)}_{L_{k+1}}(\Bx)}(E)\Bone_{\BC^{(n,int)}_{2L_k}(\Bx_{i_0})}\|\leq C_{geom}\|\Bone_{\BC^{(n,out)}_{L_{k+1}}(\Bx)}\BG^{(n)}_{\BC^{(n)}_{L_{k+1}}(\Bx)}(E)\Bone_{\BC^{(n,out)}_{L_k}(\Bx_{i_0})}\|\\
\times  \|\Bone_{\BC^{(n,out)}_{L_k}(\Bx_{i_0})}\BG^{(n)}_{\BC^{(n)}_{L_k}(\Bx_{i_0})}(E)\Bone_{\BC^{(n,int)}_{L_k}(\Bx_{i_0})}\|\\
\leq C_{geom} \ee^{(2L_k)^{1/2}}\cdot \|\Bone_{\BC^{(n,out)}_{L_{k+1}}(\Bx)}\BG^{(n)}_{\BC^{(n)}_{L_{k+1}}(\Bx)}(E)\Bone_{\BC^{(n,out)}_{2L_k}(\Bx_{i_0})}\|
\end{gather*}

We have almost everywhere
\[
\Bone_{\BC^{(n,out)}_{2L_k}(\Bx_{i_0})}\leq \sum_{\tilde{\Bx}\in\BC^{(n)}_{2L_k}(\Bx_{i_0})\cap\Gamma_{\Bx_{i_0}},\BC^{(n)}_{L_k}(\tilde{\Bx})\not\subset\BC^{(n)}_{2L_k}(\Bx_{i_0})}\Bone_{\BC^{(n,int)}_{L_k}(\tilde{\Bx})}
\]
Hence, by choosing $\tilde{\Bx}$ such that the right hand side is maximal, we get
\[
\|\Bone_{\BC^{(n,out)}_{L_{k+1}}(\Bx)}\BG^{(n)}_{\BC^{(n)}_{L_{k+1}}(\Bx)}(E)\Bone_{\BC^{(n,int)}_{2L_k}(\Bx_{i_0})}\|\leq 6^{nd}\cdot\|\Bone_{\BC^{(n,out)}_{L_{k+1}}(\Bx)}\BG^{(n)}_{\BC^{(n)}_{L_{k+1}}(\Bx)}(E)\Bone_{\BC^{(n,int)}_{L_k}(\tilde{\Bx})}\|.
\]
Since, $\BC^{(n)}_{L_k}(\tilde{\Bx})\not\subset\BC^{(n)}_{2L_k}(\Bx_{i_0})$ , $\tilde{\Bx}\in\BC^{(n)}_{2L_k}(\Bx_{i_0})$ and the cubes $\BC^{(n)}_{2L_k}(\Bx_i)$ are disjoint, we obtain that
\[
\BC^{(n)}_{L_k}(\tilde{\Bx})\not\subset \bigcup_{i=1}^r\BC^{(n)}_{2L_k}(\Bx_i),
\]
so that the cube $\BC^{(n)}_{L_k}(\tilde{\Bx})$ must be $(E,m)$-NS. We therefore perform a new step as in case (a) and obtain:

\[
\cdots \leq 6^{nd}3^{nd}C_{geom}\cdot \ee^{-\gamma(m,L_k,n)L_k}\cdot\|\Bone_{\BC^{(n,out)}_{L_{k+1}}(\Bx)}\BG^{(n)}_{\BC^{(n)}_{L_{k+1}}(\Bx)}(E)\Bone_{\BC^{(n,int)}_{L_k}(\Bx_{\ell+1})}\|,
\]

with $\Bx_{\ell+1}\in\Gamma_{\tilde{\Bx}}$ and $|\tilde{\Bx}-\Bx_{\ell+1}|=L_k/3$.
\end{enumerate}

 Summarizing, we get $\Bx_{\ell+1}$ with 

\[
\|\Bone_{\BC^{(n,out)}_{L_{k+1}}(\Bx)}\BG^{(n)}_{\BC^{(n)}_{L_{k+1}}(\Bx)}(E)\Bone_{\BC^{(n,int)}_{L_k}(\Bx_{\ell})}\|\leq \delta_{0}\cdot\|\Bone_{\BC^{(n,out)}_{L_{k+1}}(\Bx)}\BG^{(n)}_{\BC^{(n)}_{L_{k+1}}(\Bx)}(E)\Bone_{\BC^{(n,int)}_{L_{k+1}}(\Bx_{\ell+1})}\|,
\]
with $\delta_0=18^{nd} C_{geom}^2\cdot\ee^{(2L_k)^1/2}\ee^{-\gamma(m,L_k,n)L_k}$. After $\ell$ iterations  with $n_+$ steps of case (a) and $n_0$ steps of case (b), we obtain

\[
\|\Bone_{\BC^{(n,out)}_{L_{k+1}}(\Bx)}\BG^{(n)}_{\BC^{(n)}_{L_{k+1}}(\Bx)}(E)\Bone_{\BC^{(n,int)}_{L_k}(\Bx_0)}\|\leq (\delta_+)^{n_+}(\delta_0)^{n_0}\cdot\|\Bone_{\BC^{(n,out)}_{L_{k+1}}(\Bx)}\BG^{(n)}_{\BC^{(n)}_{L_{k+1}}(\Bx)}(E)\Bone_{\BC^{(n,int)}_{L_k}(\Bx_{\ell})}\|.
\]
Now since $\gamma(m,L_k,n)>m$, we have that 
\[
\delta_+\leq 3^{nd}\cdot C_{geom}\ee^{-mL_k}.
\]
So $\delta_+$ can be made arbitrarily  small if $L_0$ and hence $L_k$ is large enough. We also have for $\delta_0$:
\begin{align*}
\delta_0&=18^{nd} C_{geom}^2\ee^{(2L_k)^1/2}\ee^{-\gamma(m,n,L_k)L_k}\\
&=18^{nd} C_{geom}^2\ee^{\sqrt{2}L_k^{1/2}}\ee^{-\gamma(m,n,L_k)L_k}\\
&\leq 18^{nd} C_{geom}^2\ee^{\sqrt{2}L_k^{1/2}-mL_k}<\frac{1}{2},
\end{align*}
For large $L_0$ and hence $L_k$. Using the (GRI), we can iterate if $\BC^{(n,out)}_{L_{k+1}}(\Bx)\cap\BC^{(n)}_{L_k}(\Bx_{\ell})=\emptyset$. Thus, we can have at least $n_+$ steps of case (a) with,
\[
n_+\cdot\frac{L_k}{3}+\sum_{i=1}^r2L_k\geq \frac{L_{k+1}}{3}-\frac{L_k}{3},
\]
until the induction eventually stop. Since $r\leq J$, we can bound $n_{+}$ from below .
\begin{align*}
n_+\cdot\frac{L_k}{3}&\geq \frac{L_{k+1}}{3}-\frac{L_k}{3}-r(L_k)\\
&\geq \frac{L_{k+1}}{3}-\frac{L_k}{3}-2JL_k\\
\end{align*}
which yields
\begin{align*}
n_+&\geq \frac{L_{k+1}}{L_k}-1-6J\\
&\geq \frac{L_{k+1}}{L_k}-7J
\end{align*}
Therefore,
\begin{equation}\label{eq:NR.NS}
\|\Bone_{BC^{(n,out)}_{L_{k+1}}(\Bx)}\BG^{(n)}_{\BC^{(n)}_{L_{k+1}}(\Bx)}(E)\Bone_{\BC^{(n,int)}_{L_k}(\Bx_0)}\|\leq \delta_+^{n_+}\cdot\|\BG^{(n)}_{\BC^{(n)}_{L_{k+1}}(\Bx)}(E)\|.
\end{equation}
Finally, by $E$-nonresonance of $\BC^{(n)}_{L_{k+1}}(\Bx)$ and since we can cover  $\BC^{(n,int)}_{L_{k+1}}(\Bx)$ by $\left(\frac{L_{k+1}}{L_k}\right)^{nd}$ small cubes $\BC^{(n,int)}_{L_k}(\By)$, equation \eqref{eq:NR.NS} with $y$ instead of $\Bx_0$ yields

\begin{gather*}
\|\Bone_{\BC^{(n,out)}_{L_{k+1}}(\Bx)}\BG^{(n)}_{\BC^{(n)}_{L_{k+1}}(\Bx)}(E)\Bone_{\BC^{(n,int)}_{L_{k+1}}(\Bx)}\| \\
\leq \left(\frac{L_{k+1}}{L_k}\right)^{nd}\cdot\delta_{n_+}\cdot\ee^{L_{k+1}^{1/2}}\\
\leq \left(\frac{L_{k+1}}{L_k}\right)^{nd}\cdot\left[3^{nd}\cdot Cgeom\cdot\ee^{-\gamma(m,L_k,n)L_k}\right]^{\frac{L_{k+1}}{L_k}-7J}\ee^{L_{k+1}^{1/2}}\\
\leq L_{k+1}^{nd}L_{k+1}^{-\frac{nd}{\alpha}} C(n,d)^{\frac{L_{k+1}}{L_k}-7J}\ee^{-\gamma(m,L_k,n)(\frac{L_{k+1}}{L_k}-7J)}\times\ee^{L_{k+1}^{1/2}}\\
\leq L_{k+1}^{nd/3}\ee^{(L_{k+1}^{1/3}-7J)\ln C(n,d)}\ee^{-\gamma(m,L_k,n)(L_{k+1}^{1/3}-7J)}\ee^{L_{k+1}^{1/2}}\\
\leq \ee^{-\left[-\frac{nd}{3}\ln(L_{k+1})-L_{k+1}^{1/3}\ln(C)+7J\ln(C(n,d))+7J\ln(C(n,d))+\gamma(m,L_k,n)L_{k+1}^{1/3}-7J\gamma(m,L_k,n)-L_{k+1}^{1/2}\right]}\\
\leq \ee^{-\left[\frac{-nd}{3}\frac{\ln L_{k+1}}{L_{k+1}}-\frac{L_{k+1}^{1/3}\ln C(n,d)}+\frac{7J\ln(C(n,d))}{L_{k+1}}+\gamma(m,L_k,n)\frac{L_{k+1}^{1/3}}{L_{k+1}}-7J\frac{\gamma(m,L_k,n)}{L_{k+1}}-{L_{k+1}}^{-1/2}\right]L_{k+1}}\\
\leq \ee^{-m'L_{k+1}},
\end{gather*}
where 
\[
m'=\frac{1}{L_{k+1}}\left[ n_+\gamma(m,L_k,n)L_k-n_+\ln((2^{Nd}NdL_k^{nd-1})\right]-\frac{1}{L_{k+1}^{1/2}},
\]
with
\[
L_{k+1}L_k^{-1}-7J\leq n_+\leq L_{k+1}L_{k}^{-1}; 
\]
we obtain
\begin{align*}
m'&\geq \gamma(m,L_k,n)-\gamma(m,L_k,n)\, \frac{4JL_k}{L_{k+1}}
\\
\qquad &-\frac{1}{L_{k+1}}\frac{L_{k+1}}{L_k}\ln(2^{Nd}Nd)L_k^{nd-1})-\frac{1}{L_{k+1}^{1/2}}
\\
&\geq \gamma(m,L_k,n)-\gamma(m,L_k,n)\, 4J L_k^{-1/2}
\\
&\qquad -L_k^{-1}(\ln(2^{Nd}Nd))+(nd-1)\ln(L_k))-L_{k}^{-3/4}
\\
&\geq \gamma(m,L_k,n)[1-(4J+\ln(2^{Nd}Nd)+Nd)L_k^{-1/2}]
\end{align*}

if $L_0\geq L_2^*(J,N,d)$ for some $L^*_2(J,N,d)>0$ large enough. Since $\gamma(m,L_k,n)=m(1+L_k^{-1/8})^{N-n+1}$,
\[
\frac{\gamma(m,L_k,n)}{\gamma(m,L_{k+1},n)}=\left(\frac{1+L_k^{-1/8}}{1+L_k^{-3/16}}\right)^{N-n+1}\geq \frac{1+L_k^{-1/8}}{1+L_k^{-3/16}}
\]
Therefore we can compute
\begin{align*}
&\frac{\gamma(m,L_k,n)}{\gamma(m,L_{k+1},n)}(1-(4J+\ln(2^{Nd}Nd)+Nd)L_k^{-1/2})\\
 &\qquad \geq\frac{1+L_k^{-1/8}}{1+L_k^{-3/16}}(1-(4J+\ln(2^{Nd}Nd)+Nd)L_k^{-1/2})>1,
\end{align*}
provided $L_0\geq \tilde{L}_2^*(J,N,d)$ for some large enough $\tilde{L}_2^*(J,N,d)>0$. Finally, we obtain that $m'>\gamma(m,l_{k+1},n)$ and $|\BG_{\BC^{(n)}_{L_{k+1}}(\Bu)}(\Bu,\Bv;E)|\leq \ee^{-\gamma(m,L_{k+1},n)L_{k+1}}$. This proves the result. 
\end{proof}
The main result of this subsection is Theorem \ref{thm:fully.interactive}. We will  need the following preliminary results.

\begin{lemma}\label{lem:MD}
Given $k\geq0$, assume that property $\dsknn$ holds true for all pairs of separable FI cubes. Then for any $\ell\geq 1$
\begin{equation}\label{eq:MD}
\DP\left\{M_{\fui}(\BC_{L_{k+1}}^{(n)}(\Bu),I_0)\geq 2\ell\right\}\leq C(n,N,d,\ell)L_k^{2\ell dn\alpha}L_k^{-2\ell p\,4^{N-n}}.
\end{equation}
\end{lemma}

\begin{proof}
Suppose there exist $2\ell$ pairwise separable, fully interactive cubes $\BC_{L_k}^{(n)}(\Bu^{(j)})$ $\subset\BC_{L_{k+1}}^{(n)}(\Bu)$, $1\leq j\leq 2\ell$. Then, by Lemma \ref{lem:PI.cubes}, for any pair $\BC_{L_k}^{(n)}(\Bu^{(2i-1)})$, $\BC_{L_k}^{(n)}(\Bu^{(2i)})$, the corresponding random Hamiltonians $\BH_{\BC_{L_k}^{(n)}(\Bu^{(2i-1)})}^{(n)}$ and $\BH^{(n)}_{\BC_{L_k}^{(n)}(\Bu^{(2i)})}$ are independent, and so are their spectra and their Green functions. For $i=1,\dots,\ell$ we consider the events:
\[
\rA_i=\left\{\exists\,E\in I_0:\BC_{L_k}^{(n)}(\Bu^{(2i-1)})\text{ and $\BC_{L_k}^{(n)}(\Bu^{(2i)})$ are $(E,m_n)$-S}\right\}.
\]
Then by assumption $\dsknn$, we have, for $i=1,\dots,\ell$,
\begin{equation}
\DP\left\{\rA_i\right\}\leq L_k^{-2p\,4^{N-n}},
\end{equation}
and, by independence of events $\rA_1,\dots,\rA_{\ell}$,
\begin{equation}
\DP\Bigl\{\bigcap_{1\leq i\leq\ell}\rA_i\Bigr\}=\prod_{i=1}^{\ell}\DP(\rA_i)\leq\bigl(L_k^{-2p\,4^{N-n}}\bigr)^{\ell}.
\end{equation}
To complete the proof, note that the total number of different families of $2\ell$ cubes $\BC_{L_k}^{(n)}(\Bu^{(j)})\subset\BC_{L_{k+1}}^{(n)}(\Bu)$, $1\leq j\leq 2\ell$, is bounded by 
\[
\frac{1}{(2\ell)!}\left|\BC_{L_{k+1}}^{(n)}(\Bu)\right|^{2\ell}\leq C(n,N,\ell,d)L_{k}^{2\ell dn\alpha}.\qedhere
\]
\end{proof}

\begin{theorem}\label{thm:fully.interactive}
Let $1\leq n\leq N$. There exists $L_2^*=L_2^*(N,d)>0$ such that if $L_0\geq L_2^*$ and if for $k\geq 0$
\begin{enumerate}[\rm(i)]
\item
$\dsn{k-1,n'}$ for all $1\leq n'<n$ holds true,
\item
$\dsn{k,n}$ holds true for all pairs of FI cubes,
\end{enumerate}
then $\dskonn$ holds true
for any pair of separable FI cubes $\BC_{L_{k+1}}^{(n)}(\Bx)$ and $\BC_{L_{k+1}}^{(n)}(\By)$.
\end{theorem}

Above, we used the convention that $\dsn{-1,n}$ means no assumption.

\begin{proof}
Consider a pair of separable FI cubes $\BC_{L_{k+1}}^{(n)}(\Bx)$, $\BC_{L_{k+1}}^{(n)}(\By)$ and set $J=\kappa(n)+5$. Define
\begin{align*}
\rB_{k+1}
&=\left\{\exists\,E\in I_0:\BC_{L_{k+1}}^{(n)}(\Bx)\text{ and $\BC_{L_{k+1}}^{(n)}(\By)$ are $(E,m)$-S}\right\},\\
\Sigma
&=\left\{\exists\,E\in I_0:\text{neither $\BC_{L_{k+1}}^{(n)}(\Bx)$ nor $\BC_{L_{k+1}}^{(n)}(\By)$ is $E$-CNR}\right\},\\
\rS_{\Bx}
&=\left\{\exists\,E\in I_0:M(\BC_{L_{k+1}}^{(n)}(\Bx);E)\geq J+1\right\},\\
\rS_{\By}
&=\left\{\exists\,E\in I_0:M(\BC_{L_{k+1}}^{(n)}(\By),E)\geq J+1\right\}.
\end{align*}
Let $\omega\in\rB_{k+1}$. If $\omega\notin\Sigma\cup\rS_{\Bx}$, then $\forall E\in I_0$ either $\BC_{L_{k+1}}^{(n)}(\Bx)$ or $\BC_{L_{k+1}}^{(n)}(\By)$ is $E$-CNR and $M(\BC_{L_{k+1}}^{(n)}(\Bx),E)\leq J$. The cube $\BC_{L_{k+1}}^{(n)}(\Bx)$ cannot be $E$-CNR: indeed, by Lemma \ref{lem:CNR.NS} it would be $(E,m)$-NS. So the cube $\BC_{L_{k+1}}^{(n)}(\By)$ is $E$-CNR and $(E,m)$-S. This implies again by Lemma \ref{lem:CNR.NS} that
\[
M(\BC_{L_{k+1}}^{(n)}(\By),E)\geq J+1.
\]
Therefore $\omega\in\rS_{\By}$, so that $\rB_{k+1}\subset\Sigma\cup\rS_{\Bx}\cup\rS_{\By}$, hence
\[
\DP\left\{\rB_{k+1}\right\}
\leq\DP\{\Sigma\}+\DP\{\rS_{\Bx}\}+\DP\{\rS_{\By}\}
\]
and $\prob{\Sigma}\leq L_{k+1}^{-4^N\,p}$ by Theorem \ref{thm:Wegner}.
Now let us estimate $\DP\{\rS_{\Bx}\}$ and similarly $\DP\{\rS_{\By}\}$. Since 
\[
M_{\pai}(\BC_{L_{k+1}}^{(n)}(\Bx),E)+M_{\fui}(\BC_{L_{k+1}}^{(n)}(\Bx),E)\geq M(\BC_{L_{k+1}}^{(n)}(\Bx),E),
\] 
the inequality $M(\BC_{L_{k+1}}^{(n)}(\Bx),E)\geq\kappa(n)+ 6$, implies that either $M_{\pai}(\BC_{L_{k+1}}^{(n)}(\Bx),E)\geq\kappa(n)+ 2$, or $M_{\fui}(\BC_{L_{k+1}}^{(n)}(\Bx),E)\geq 4$. Therefore, by Lemma \ref{lem:MND} and Lemma \ref{lem:MD} (with $\ell=2$),
\begin{align*}
\DP\{\rS_{\Bx}\}&\leq\DP\left\{\exists\,E\in I:M_{\pai}(\BC_{L_{k+1}}^{(n)}(\Bx),E)\geq \kappa(n)+2\right\}\\
&\quad+\DP\left\{\exists\,E\in I:M_{\fui}(\BC_{L_{k+1}}^{(n)}(\Bx),E)\geq 4\right\}\\
&\leq\frac{3^{2nd}}{2}L_{k+1}^{2nd}(L_k^{-4^Np}+L_k^{-4p\,4^{N-n}})+C'(n,N,d)L_{k+1}^{4 dn-\frac{4 p}{\alpha}4^{N-n}}\\
&\leq C''(n,N,d)\left(L_{k+1}^{-\frac{4^Np}{\alpha}+2nd}+L_{k+1}^{-\frac{4p}{\alpha}4^{N-n}+2nd}+L_{k+1}^{-\frac{4p}{\alpha}4^{N-n}+4nd}\right)\\
&\leq C'''(n,N,d) L_{k+1}^{-\frac{4p}{\alpha}4^{N-n}+4nd}\tag{$\alpha=3/2$}\\
&\leq\frac{1}{4}L_{k+1}^{-2p\,4^{N-n}},
\end{align*}
where we used $p>4\alpha Nd=6Nd$. Finally 
\[ 
\prob{\rB_{k+1}}\leq L_{k+1}^{-4^Np}+\frac{1}{2}L_{k+1}^{-2p4^{N-n}}<L_{k+1}^{-2p4^{N-n}}.
\]
\end{proof}

\subsection{Mixed pairs of cubes}\label{ssec:mixed.S}
Finally, it remains only to derive $\dskonn$ in case (III), i.e., for pairs of $n$-particle cubes where one is PI while the other is FI.

\begin{theorem}\label{thm:mixed}
Let $1\leq n\leq N$. There exists $L_3^*=L_3^*(N,d)>0$ such that if $L_0\geq L_3^*(N,d)$ and if for $k\geq 0$,
\begin{enumerate}[\rm(i)]
\item
$\dsn{k-1,n'}$ holds true for all $1\leq n'<n$,
\item
$\dsn{k,n'}$ holds true for all $1\leq n'<n$ and
\item
$\dsknn$ holds true for all pairs of FI cubes,
\end{enumerate}
then $\dskonn$ holds true for any pair of separable cubes $\BC_{L_{k+1}}^{(n)}(\Bx)$, $\BC_{L_{k+1}}^{(n)}(\By)$ where one is PI while the other is FI.
\end{theorem}

\begin{proof}
Consider a pair of separable $n$-particle cubes $\BC_{L_{k+1}}^{(n)}(\Bx)$, $\BC_{L_{k+1}}^{(n)}(\By)$ and suppose that $\BC_{L_{k+1}}^{(n)}(\Bx)$ is PI while $\BC_{L_{k+1}}^{(n)}(\By)$ is FI. Set $J=\kappa(n)+5$ and introduce the events
\begin{align*}
\rB_{k+1}
&=\left\{\exists\,E\in I_0:\BC_{L_{k+1}}^{(n)}(\Bx)\text{ and $\BC_{L_{k+1}}^{(n)}(\By)$ are $(E,m)$-S}\right\},\\
\Sigma
&=\left\{\exists\,E\in I_0: \BC_{L_{k+1}}^{(n)}(\Bx)\ \text{is not $E$-CNR and}\ \BC_{L_{k+1}}^{(n)}(\By)\text{ is not $E$-CNR}\right\},\\
\mathcal{N}_{\Bx}
&=\left\{ \BC_{L_{k+1}}^{(n)}(\Bx)\text{ is $m$-non-localized}\right\},\\
\rS_{\By}
&=\left\{ \exists\,E\in I_0:M(\BC_{L_{k+1}}^{(n)}(\By),E)\geq J+1\right\}.
\end{align*}
Let $\omega\in \rB_{k+1}\setminus (\Sigma\cup \mathcal{N}_{\Bx})$, then for all $E\in I_0$ either $\BC_{L_{k+1}}^{(n)}(\Bx)$ is $E$-CNR or $\BC_{L_{k+1}}^{(n)}(\By)$ is $E$-CNR and  $\BC_{L_{k+1}}^{(n)}(\Bx)$ is $m$-localized. The cube $\BC_{L_{k+1}}^{(n)}(\Bx)$ cannot be $E$-CNR. Indeed, by Lemma \ref{lem:localized} it would have been $(E,m)$-NS. Thus the cube $\BC_{L_{k+1}}^{(n)}(\By)$ is $E$-CNR, so by Lemma \ref{lem:CNR.NS}, $M(\BC_{L_{k+1}}^{(n)}(\By);E)\geq J+1$: otherwise $\BC_{L_{k+1}}^{(n)}(\By)$ would be $(E,m)$-NS. Therefore $\omega\in \rS_{\By}$. Consequently,
\[
\rB_{k+1}\subset \Sigma \cup \mathcal{N}_{\Bx}\cup\rS_{\By}.
\]
Recall that the probabilities $\DP\{\mathcal{N}_{\Bx}\}$ and $\DP\{\rS_{\By}\}$ have already been estimated in Sections \ref{ssec:PI.cubes} and \ref{ssec:FI.cubes}. We therefore obtain
\begin{align*}
\DP\left\{\rB_{k+1}\right\}&\leq \DP\{\Sigma\}+\DP\{\mathcal{N}_{\Bx}\}+\DP\{\rS_{\By}\}\\
&\leq L_{k+1}^{-4^Np}+ \frac{1}{2}L_{k+1}^{-4p\,4^{N-n}}+ \frac{1}{4}L_{k+1}^{-2p\,4^{N-n}}\leq L_{k+1}^{-2p\,4^{N-n}}.\qedhere
\end{align*}
\end{proof}

\section{Conclusion: the multi-particle multi-scale analysis}

\begin{theorem}\label{thm:DS.k.N}
Let $1\leq n\leq N$ and  $\BH^{(n)}(\omega)=-\BDelta+\sum_{j=1}^n V(x_j,\omega)+\BU$, where $\BU$, $V$ satisfy
 $\condI$ and $\condP$  respectively. There exists $m_n>0$ such that for any $p>6Nd$ property $\dsknn$ holds true for all $k\geq 0$ provided $L_0$ is large enough.
\end{theorem}
\begin{proof}
We prove that for each $n=1,\ldots,N$, property $\dsknn$ is valid. To do so, we use an induction on the number of particles $n'=1,\ldots,n$. For $n=1$  property $\dsn{k,1}$ holds true for all $k\geq 0$ by  the single-particle localization theory \cite{St01}. Now suppose that for all $1\leq n'<n$, $\dsn{k,n'}$ holds true for all	 $k\geq 0$, we aim to prove that $\dsknn$ holds true for all $k\geq 0$. For $k=0$, $\dsn{0,n}$ is valid using Theorem \ref{thm:initial.var.energy}. Next, suppose that $\dsn{k',n}$ holds true for all $k'<k$, then by  combining this last assumption with $\dsn{k,n'}$ above, one can conclude that
\begin{enumerate}
\item[\rm(i)] $\dsknn$ holds true for all $k\geq 0$ and for all pairs of PI cubes using Theorem \ref{thm:partially.interactive},
\item[\rm(ii)] $\dsknn$ holds true for all $k\geq 0$ and for all pairs of FI cubes using Theorem \ref{thm:fully.interactive},
\item[\rm(iii)] $\dsknn$ holds true for all $k\geq 0$ and for all pairs of MI cubes using Theorem \ref{thm:mixed}.
\end{enumerate}
 Hence Theorem \ref{thm:DS.k.N} is proven.
\end{proof}

\section{Proofs of the results}

\subsection{Proof of Theorem \ref{thm:exp.loc}}
		
Using the multi-particle multi-scale analysis bounds in the continuum property $\dsn{k,N}$, we extend to multi-particle systems the strategy of Stollmann \cite{St01}.	

For $\Bx_0\in\DZ^{Nd}$ and an integer $k\geq 0$, set, using the notations of Lemma \ref{lem:separable.distant}
\[
R(\Bx_0):=\max_{1\leq \ell\leq \kappa(N)}|\Bx_0-\Bx^{(\ell)}|;\qquad  b_{k}(\Bx_0):=7N+R(\Bx_0)L_k^{-1},
\]
\[
M_k(\Bx_0):=\bigcup_{\ell=1}^{\kappa(N)}\BC^{(N)}_{7NL_k}(\Bx^{(\ell)}
\] 
and define
\[
A_{k+1}(\Bx_0):=\BC^{(N)}_{bb_{k+1}L_{k+1}}(\Bx_0)\setminus \BC^{(N)}_{b_kL_k}(\Bx_0),
\]
where the parameter $b>0$ is to be chosen later. We can easily check that,
\[
M_k(\Bx_0)\subset \BC^{(N)}_{b_kL_k}(\Bx_0).
\]
Moreover, if $\Bx\in A_{k+1}(\Bx_0)$, then the cubes $\BC^{(N)}_{L_k}(\Bx)$ and $\BC^{(N)}_{L_k}(\Bx_0)$ are separable by Lemma \ref{lem:separable.distant}. Now, also define
\[
\Omega_k(\Bx_0):=\{\text{$\exists E\in I_0$ and $\Bx\in A_{k+1}(\Bx_0)\cap \Gamma_k$: $\BC^{(N)}_{L_k}(\Bx)$ and $\BC^{(N)}_{L_k}(\Bx_0)$ are $(E,m)$-S}\},
\]
with $\Gamma_k:= \Bx_0+\frac{L_k}{3}\DZ^{Nd}$. Now, property $\dsn{k,N}$ combined with the cardinality of $A_{k+1}(\Bx_0)\cap \Gamma_k$ imply 
\begin{align*}
\prob{\Omega_k(\Bx_0)}&\leq (2bb_{k+1}L_{k+1})^{Nd}L_k^{-2p}\\
&\leq (2bb_{k+1})^{Nd}L^{-2p+\alpha Nd}. 
\end{align*}
Since, $p>(\alpha Nd+1)/2$ (in fact, $p>6Nd$), we get 
\[
\sum_{k=0}^{\infty}\prob{\Omega_k(\Bx_0)}<\infty.
\]
Thus, setting 
\[
\Omega_{<\infty}:=\{\text{ $\forall \Bx_0\in\DZ^{Nd}$, $\Omega_k(\Bx_0)$ occurs finitely many times} \},
\]
by the Borel cantelli Lemma and the countability of $\DZ^{Nd}$ we have that $\prob{\Omega_{<\infty}}=1$. Therefore it suffices to pick $\omega\in \Omega_{<\infty}$ and prove the exponential decay of any nonzero eigenfunction $\BPsi$ of $\BH^{(N)}(\omega)$.

Let $\BPsi$ be a polynomially bounded  eigenfunction satisfying (EDI) (see Theorem \ref{thm:GRI.EF}). Let $\Bx_0\in\DZ^{Nd}$ with $\|\Bone_{\BC^{(N)}_1(\Bx_0)}\BPsi\|>0$ (if there is no such $\Bx_0$, we are done). The cube $\BC^{(N)}_{L_k}(\Bx_0)$ cannot be $(E,m)$-NS for infinitely many $k$. Indeed, given an integer $k\geq0$, if $\BC^{(N)}_{L_k}(\Bx_0)$ is $(E,m)$-NS  then by (EDI), and the polynomial bound on $\BPsi$, we get
\begin{align*}
\|\Bone_{\BC^{(N)}_1(\Bx_0)}\BPsi\|&\leq C\cdot\|\Bone_{\BC^{(N,out)}_{L_k}(\Bx_0)}\BG^{(N)}_{\BC^{(N)}_{L_k}(\Bx_0}(E)\Bone_{\BC^{(N,int)}_{L_k}(\Bx_0}\|\cdot\|\Bone_{\BC^{(N,out)}_{L_k}(\Bx_0)}\BPsi\|\\
&\leq C(1+|\Bx_0|+L_k)^t\cdot \ee^{-mL_k}\tto{L_k\rightarrow \infty} 0,
\end{align*}
in contradiction with the choice of $\Bx_0$. So there is an integer $k_1=k_1((\omega,E,\Bx_0)<\infty$ such that $\forall k\geq k_1$ the cube $\BC^{(N)}_{L_k}(\Bx_0)$ is $(E,m)$-S.  At the same time, since $\omega\in \Omega_{<\infty}$, there exists $k_2=k_2(\omega,\Bx_0)$ such that if $k\geq k_2$, $\Omega_k(\Bx_0)$ does not occurs. We conclude that for all $k\geq \max\{k_1,k_2\}$, for all $\Bx\in A_{k+1}(\Bx_0)\cap\Gamma_k$, $\BC^{(N)}_{L_k}(\Bx)$ is $(E,m)$-NS. 

Let $\rho\in(0,1)$ and choose $b>0$ such that
 \[
b>\frac{1+\rho}{1-\rho},
\]
so that
\[
\BC^{(N)}_{\frac{bb_{k+1}L_{k+1}}{1-\rho}}(\Bx_0)\setminus\BC^{(N)}_{\frac{b_kL_k}{1-\rho}}(\Bx_0)\subset A_{k+1}(\Bx_0),
\]
for $\Bx\in\tilde{A}_{k+1}(\Bx_0)$.
\begin{enumerate}
\item[(1)]
Since, $|\Bx-\Bx_0|>\frac{b_kL_k}{1-\rho}$,
\begin{align*}
\dist(\Bx,\partial\BC^{(N)}_{b_kL_k}(\Bx_0)&\geq |\Bx-\Bx_0|-b_kL_k\\
&\geq |\Bx-\Bx_0|-(1-\rho)|\Bx-\Bx_0|\\
&=\rho(|\Bx-\Bx_0|)
\end{align*}                              

\item[(2)]
Since $|\Bx-\Bx_0|\leq \frac{bb_{k+1}L_{k+1}}{1+\rho}$,
\begin{align*}
\dist(\Bx,\partial\BC^{(N)}_{bb_{k+1}L_{k+1}}(\Bx_0))&\geq bb_{k+1}L_{k+1}-|\Bx-\Bx_0|\\
&\geq (1+\rho)|\Bx-\Bx_0|-|\Bx-\Bx_0|\\
&=\rho|\Bx-\Bx_0|.
\end{align*}
\end{enumerate}
Thus, 
\[
\dist(\Bx,\partial A_{k+1}(\Bx_0))\geq \rho|\Bx-\Bx_0|.
\]
Now, setting $k_3=\max\{k_1,k_2\}$, the assumption linking $b$ and $\rho$ implies that:
\[
\bigcup_{k\geq k_3} \tilde{A}_{k+1}(\Bx_0)=\DR^{Nd}\setminus\BC^{(N)}_{\frac{b_{k_3}L_{k_3}}{1-\rho}}(\Bx_0),
\]
because $\frac{bb_{k+1}L_{k+1}}{1+\rho}>\frac{b_kL_k}{1-\rho}$. Let $k\geq k_3$, recall that this implis that all the cubes with centers in $A_{k+1}(\Bx_0)\cap\Gamma_k$ and side length $2L_k$ are $(E,m)$-NS. Thus, for any $\Bx\in\tilde{A}_{k+1}(\Bx_0)$, we choose $\Bx_1\in A_{k+1}(\Bx_0)$ such that $\Bx\in \BC^{(N,int)}_{L_k}(\Bx_1)$. Therefore
\begin{align*}
\|\Bone_{\BC^{(N)}_1(\Bx)}\BPsi\|&\leq \|\Bone_{\BC^{(N,int)}_{L_k}(\Bx_1)}\BPsi\|\\
&\leq C\cdot\ee^{-mL_k}\cdot\|\Bone_{\BC^{(N,out)}_{L_k}(\Bx_1)}\BPsi\|
\end{align*}

Up to a set of Lebesgue measure zero, we can cover $\BC^{(N,out)}_{L_k}(\Bx_1)$ by at most $3^{Nd}$ cubes 
\[
\BC^{(N,int)}_{L_k}(\tilde{\Bx}),\qquad \tilde{\Bx}\in\Gamma_k,\quad |\tilde{\Bx}-\Bx_1|=\frac{L_k}{3}.
\]
By choosing $\Bx_2$ which gives a maximal norm, we get
\[
\|\Bone_{\BC^{(N,out)}_{L_k}(\Bx_1)}\BPsi\|\leq 3^{Nd}\cdot\|\Bone_{\BC^{(N,int)}_{L_k}(\Bx_2)}\BPsi\|,
\]
so that 
\[
\|\Bone_{\BC^{(N)}_1(\Bx)}\BPsi\|\leq  3^{Nd}\cdot\ee^{-m L_k}\cdot\|\Bone_{\BC^{(N,int)}_{L_k}(\Bx_2)}\BPsi\|.
\]
Thus, by an induction procedure, we find a sequence $\Bx_1$, $\Bx_2$, ..., $\Bx_n$ in $\Gamma_k\cap A_{k+1}(\Bx_0)$  and the bound
\[
\|\Bone_{\BC^{(N)}_1(\Bx)}\BPsi\|\leq (C\cdot 3^{Nd}\exp(-mL_k))^n\cdot\|\Bone{\BC^{(N,out)}_{L_k}(\Bx_n)}\BPsi\|.
\]
Since $|\Bx_i-\Bx_{i+1}|=L_k/3$ and  $\dist(\Bx,\partial A_{k+1})\geq \rho\cdot|\Bx-\Bx_0|$, we can iterate at least $\rho\cdot|\Bx-\Bx_0|\cdot3/L_k$ times until, we reach the boundary of $A_{k+1}(\Bx_0)$. Next, using the polynomial bound on $\BPsi$, we obtain:

\begin{gather*}
\|\Bone_{\BC^{(N)}_1(\Bx)}\BPsi\|\leq (C\cdot 3^{Nd})^{\frac{3\rho|\Bx-\Bx_0|}{L_k}}\cdot\exp(-3m\rho|\Bx-\Bx_0|)\\
\times C(1+|\Bx_0|+ bL_{k+1})^t\cdot L_{k+1}^{Nd}.
\end{gather*}

We can conclude that given $\rho'$ with $0<\rho'<1$, we can find $k_4\geq k_3$ such that if $k\geq k_4$, then
\[
\|\Bone_{\BC^{(N)}_1(\Bx)}\BPsi\|\leq \ee^{-\rho\rho'm|\Bx-\Bx_0|},
\]
if $|\Bx-\Bx_0|>\frac{b_{k_4}L_{k_4}}{1-\rho}$. This completes the proof of the exponential localization in the max-norm.

\subsection{Proof of Theorem \ref{thm:dynamical.loc}}
		
For the proof of the multi-particle dynamical localization given  the multi-particle multi-scale analysis in the continuum, we refer to the paper by Boutet de Monvel et al. \cite{BCS11}.

\section{Appendix}\label{sec:appendix}
\begin{center}

\end{center}

\subsection{Proof of Lemma \ref{lem:separable.distant}}

(A) Let $L>0$, $\emptyset \neq \CJ\subset \{1,\ldots,n\}$ and $\By\in\DZ^{nd}$. $\{y_j\}_{j\in\CJ}$ is called an $L$-cluster if the union
\[
\bigcup_{j\in\CJ} C^{(1)}_L(y_j)
\]
cannot be decomposed into two non-empty disjoint subsets. Next, given two configurations $\Bx,\By\in\DZ^{nd}$, we proceed as follows:
\begin{enumerate}
\item
We decompose the vector $\By$ into maximal $L$-clusters $\Gamma_1,\ldots,\Gamma_M$ (each of diameter $\leq 2nL$) with $M\leq n$.
\item
Each position $y_i$ corresponds to exactly one cluster $\Gamma_j$, $j=j(i)\in\{1,\ldots,M\}$.
\item
If there exists $j\in\{1,\ldots,M\}$ such that $\Gamma_j\cap\varPi\BC^{(n)}_L(\Bx)=\emptyset$, then cubes $\BC^{(n)}_L(\By)$ and $\BC^{(n)}_L(\Bx)$ are separable.
\item
If (3)  is wrong, then for all $k=1,\ldots,M$, $\Gamma_k\cap \varPi\BC^{(n)}_L(\Bx)\neq \emptyset$. Thus for all $k=1,\ldots,M$, $\exists i=1,\ldots,n$ such that $\Gamma_k\cap \BC^{(1)}_L(x_i)\neq \emptyset$. Now for any $j=1,\ldots,n$ there exists $k=1,\ldots,M$ such that $y_j\in\Gamma_k$. Therefore for such $k$, by hypothesis there exists $i=1,\ldots,n$ such that $\Gamma_k\cap C^{(1)}_L(x_i)\neq \emptyset$. Next let $z\in\Gamma_k\cap C^{(1)}_L(x_i)$ so that $|z-x_i|\leq L$. We have that
\begin{align*}
|y_j-x_i|&\leq |y_j-z|+|z-x_i|\\
&\leq 2nL-L + L=2nL
\end{align*}
\end{enumerate}
since $y_j,z\in\Gamma_k$. Notice that above we have the bound $|y_j-z|\leq 2nL-L$ instead of $2nL$ because $y_j$ is a center of the $L$-cluster $\Gamma_k$. Hence for all $j=1,\ldots,n$ $y_j$ must belong to one of the cubes $C^{(1)}_{2nL}(x_i)$ for the $n$ positions $(y_1,\ldots,y_n)$. Set $\kappa(n)=n^n$. For any choice of at most $\kappa(n)$ possibilities, $\By=(y_1,\ldots,y_n)$ must belong to the Cartesian product of $n$ cubes of size $2nL$ i.e. to an $nd$-dimensional  cube of size  $2nL$, the assertion then follows.

(B) Set $R(\By)=\max_{1\leq i,j\leq n}|y_i-y_j|+5NL$ and consider a cube $\BC^{(n)}_L(\Bx)$ with $|\By-\Bx|> R(\By)$. Then there exists $i_0\in\{1,\ldots,n\}$ such that $|y_{i_0}-x_{i_0}|>R(\By)$. Consider the maximal connected component $\Lambda_{\Bx}:=\bigcup_{i\in\CJ} C^{(1)}_{L}(x_i)$ of the union $\bigcup_i C^{(1)}_L(x_i)$ containing $x_{i_0}$. Its diameter is bounded by $2nL$. We have
\[
\dist(\Lambda_{\Bx},\varPi\BC^{(n)}_L(\By))=\min_{u,v}|u-v|,
\]
now since
\[
|x_{i_0}-y_{i_0}|\leq |x_{i_0}-u|+|u-v|+|v-y_{i_0}|,
\]
then

\begin{align*}
\dist(\Lambda_{\Bx},\varPi\BC^{(n)}_L(\By))&=\min_{u,v}|u-v|\\
&\geq |x_{i_0}-y_{i_0}|-\diam(\Lambda_{\Bx})-\max_{v,y_{i_0}}|v-y_{i_0}|.\\
\end{align*}
Recall that $\diam(\Lambda_{\Bx})\leq 2nL$ and
\[
\max_{v,y_{i_0}}|v-y_{i_0}|\leq \max_{v}|v-y_j|+\max_{y_{i_0}}|y_j-y_{i_0}|,
\]
for some $j=1,\cdots n $ such that $v\in C^{(1)}_L(y_j)$. Finally we get
\[
\dist(\Lambda_{\Bx},\varPi\BC^{(n)}_L(\By))>R(\By) -\diam(\Lambda_{\Bx})-(2L + \diam(\varPi\By))>0,
\]
this implies that $\BC^{(n)}_L(\Bx)$ is $\CJ$-separable from $\BC^{(n)}_L(\By)$ with $\CJ$ the index subset appearing in the definition of $\Lambda_{\Bx}$.

\subsection{Proof of Lemma \ref{lem:PI.cubes}}
It is convenient to use the canonical injection $\DZ^d\hookrightarrow\DR^d$; then the notion of connectedness in $\DR^d$ induces its analog for lattice cubes. Set $R:=2L+r_0$ and assume that $\diam\varPi\Bu = \max_{i,j}|u_i - u_j|>nR$.
If the union of cubes $C_{R/2}^{(1)}(u_i)$, $1\leq i\leq n$, were not decomposable into two (or more) disjoint groups, then it would be connected, hence its diameter would be bounded by $n(2(R/2))=nR$, hence $\diam \varPi\Bu\leq nR$ which contradicts the hypothesis. Therefore, there exists an index subset $\CJ\subset\{1,\ldots,n\}$ such that $|u_{j_1}-u_{j_2}|>2(R/2)$ for all $j_1\in\CJ$, $j_2\in\CJ^c$, this implies that
\begin{align*}
\dist\left(\varPi_{\CJ}\BC_L^{(n)}(\Bu),\varPi_{\CJ^{\comp}}\BC_L^{(n)}(\Bu)\right)&=\min_{j_1\in\CJ,j_2\in\CJ^c}\dist\left(C^{(1)}_L(u_{j_1}),C^{(1)}_L(u_{j_2})\right)\\
&\geq \min_{j_1\in\CJ,j_2\in\CJ^{c}}|u_{j_1}-u_{j_2}|-2L>r_0.
\end{align*}

\subsection{Proof of Lemma \ref{lem:FI.cubes}}

If for some $R>0$,
\[
R<|\Bx-\By|=\max_{1\leq j\leq n}|x_j-y_j|,
\]
then there exists $1\leq j_0\leq n$ such that $|x_{j_0}-y_{j_0}|>R$. Since both cubes are fully interactive, by Definition \eqref{def:diagonal.cubes}
\begin{align*}
&|x_{j_0}-x_i| \le \diam \varPi \Bx \le n(2L+r_0),\\
&|y_{j_0}-y_j| \le \diam \varPi \By \le n(2L+r_0).
\end{align*}
By the triangle inequality, for any $1\leq i,j\leq n$ and $R>7\,nL >6\,nL+2\,nr_0$, we have
\begin{align*}
|x_i-y_j|&\geq |x_{j_0}-y_{j_0}|-|x_{j_0}-x_i|-|y_{j_0}-y_j|\\
&>6nL+2nr_0-2n(2L+r_0)=2nL.
\end{align*}
Therefore, for any $1\leq i,j\leq n$,
\[
\min_{i,j}\dist(C^{(1)}_L(x_i),C^{(1)}_L(y_j))\geq \min_{i,j}|x_i-y_j|-2L>2(n-1)L\geq 0,
\]
which proves the claim.

\subsection{Proof of Lemma \ref{lem:MPI}}
Assume that $M^{\sep}(\BC^{(n)}_{L_{k+1}}(\Bu),E)<2$, (i.e., there is no pair of separable  cubes of radius $L_k$ in $\BC^{(n)}_{L_{k+1}}(\Bu)$), but $M(\BC^{(n)}_{L_{k+1}}(\Bu),E)\geq \kappa(n)+2$. Then $\BC^{(n)}_{L_{k+1}}(\Bu)$ must contain at least $\kappa(n)+2$ cubes $\BC^{(n)}_{L_k}(\Bv_i)$, $0\leq i\leq \kappa(n)+1$ which are non separable but satisfy $|\Bv_i-\Bv_{i'}|>7NL_k$, for all $i\neq i'$. On the other hand, by Lemma \ref{lem:separable.distant} there are at most $\kappa(n)$ cubes $\BC^{(n)}_{2nL_k}(\By_i)$, such that any cube $\BC^{(n)}_{L_k}(\Bx)$ with $\Bx\notin \bigcup_{j} \BC^{(n)}_{2nL_k}(\By_j)$ is separable from $\BC^{(n)}_{L_k}(\Bv_0)$. Hence $\Bv_i\in \bigcup_{j} \BC^{(n)}_{2nL_k}(\By_j)$ for all $i=1,\ldots,\kappa(n)+1$.  But since for all $i\neq i'$, $|\Bv_i-\Bv_{i'}|>7NL_k$, there must be at most one center $\Bv_i$ per cube $\BC^{(n)}_{2nL_k}(\By_j)$, $1\leq j\leq \kappa(n)$. Hence we come to a contradiction:
\[
\kappa(n)+1\leq \kappa(n).
\]
 The same analysis holds true if we consider only PI cubes.

\subsection{Proof of Lemma \ref{lem:MND}}
Suppose that $M_{\pai}(\BC^{(n)}_{L_{k+1}}(\Bu),I)\geq \kappa(n)+2$, then by Lemma \ref{lem:MPI}, $M_{\pai}^{\sep}(\BC^{(n)}_{L_{k+1}}(\Bu), I)\geq 2$, i.e., there are at least two separable $(E,m)$-singular PI cubes $\BC^{(n)}_{L_k}(\Bu^{(j_1)})$, $\BC^{(n)}_{L_k}(\Bu^{(j_2)})$ inside $\BC^{(n)}_{L_{k+1}}(\Bu)$.
The  number of possible pairs of centers $\{\Bu^{(j_1)},\Bu^{(j_2)}\}$ such that
\[
\BC_{L_k}^{(n)}(\Bu^{(j_1)}),\,\BC_{L_k}^{(n)}(\Bu^{(j_2)})\subset\BC_{L_{k+1}}^{(n)}(\Bu)
\]
 is bounded by $\frac{3^{2nd}}{2}L_{k+1}^{2nd}$. Then, setting 
\[
\rB_k=\{\text{$\exists E\in I$, $\BC^{(n)}_{L_k}(\Bu^{(j_1)})$, $\BC^{(n)}_{L_k}(\Bu^{(j_2)})$ are $(E,m)$-S}\},
\]
\[
\DP\left\{M_{\pai}^{\sep}(\BC_{L_{k+1}}^{(n)}(\Bu),I)\geq 2\right\}\leq\frac{3^{2nd}}{2}L_{k+1}^{2nd}\times\prob{\rB_k}
\]
with  $\prob{\rB_k}\leq L_k^{-4^Np}+L_k^{-4p\,4^{N-n}}$ by \eqref{eq:bound.PI}.
Here $\rB_k$ is defined as in Theorem \ref{thm:partially.interactive}.

\end{document}